\documentclass[conference]{IEEEtran}
\IEEEoverridecommandlockouts
\usepackage{cite}
\usepackage{amsmath,amssymb,amsfonts}
\usepackage{graphicx}
\usepackage{textcomp}
\usepackage{xcolor}
\usepackage[linesnumbered,ruled,vlined]{algorithm2e}
\usepackage{balance}
\usepackage{multirow}
\usepackage{enumitem}
\usepackage{soul}
\usepackage{tikz}
\usepackage{calc}
\usepackage{makecell}
\usepackage{amsmath}
\usepackage{booktabs}
\usepackage{subfigure}
\usepackage{tikz}
\usepackage{framed}
\usepackage{lipsum}
\usepackage{color}
\definecolor{shadecolor}{rgb}{0.92,0.92,0.92}

\newcommand*\circled[1]{\tikz[baseline=(char.base)]{
            \node[shape=circle,fill,inner sep=0.01pt] (char) {\textcolor{white}{#1}};}}
\newcommand*\scircled[1]{\tikz[baseline=(char.base)]{
            \node[shape=circle,fill,inner sep=1.15pt] (char) {\textcolor{white}{#1}};}}
\newtheorem{definition}{Definition}

\def\BibTeX{{\rm B\kern-.05em{\sc i\kern-.025em b}\kern-.08em
    T\kern-.1667em\lower.7ex\hbox{E}\kern-.125emX}}
\begin{document}

\NewDocumentCommand{\cc}{ mO{} }{\textcolor{blue}{\textsuperscript{\textit{CC}}\textsf{\textbf{\small[#1]}}}}
 
\newcommand{\lgl}[1]{\textcolor{blue}{#1}}
\newcommand{\chao}[1]{{\textcolor{blue}{#1}}}
\NewDocumentCommand{\jt}{ mO{} }{\textcolor{magenta}{\textsuperscript{\textit{Jt}}\textsf{\textbf{\small[#1]}}}}
\newcommand{\oursys}{\ensuremath{\tt AutoCE}\xspace}
\newcommand{\autoce}{\ensuremath{\tt AutoCE}\xspace}
\newcommand{\withoutcl}{\texttt{WithoutCL}\xspace}
\newcommand{\feedback}{\texttt{Feedback}\xspace}
\newcommand{\nofeedback}{\texttt{NonFeedback}\xspace}
\newcommand{\opt}{\texttt{Optimum}\xspace}
\newcommand{\ruleb}{\texttt{Rule}\xspace}
\newcommand{\mlpb}{\texttt{MLP}\xspace}
\newcommand{\sampling}{\texttt{Sampling}\xspace}
\newcommand{\knn}{\texttt{Knn}\xspace}
\newcommand{\regr}{\texttt{Rg-based}\xspace}
\newcommand{\glayer}{\texttt{GINConV}\xspace}
\newcommand{\bayes}{\texttt{BayesCard}\xspace}
\newcommand{\deepdb}{\texttt{DeepDB}\xspace}
\newcommand{\neuro}{\texttt{NeuroCard}\xspace}
\newcommand{\uae}{\texttt{UAE}\xspace}
\newcommand{\mscn}{\texttt{MSCN}\xspace}
\newcommand{\xgb}{\texttt{LW-XGB}\xspace}
\newcommand{\nn}{\texttt{LW-NN}\xspace}
\newcommand{\imdb}{\texttt{IMDB-20}\xspace}
\newcommand{\stats}{\texttt{STATS-20}\xspace}
\newtheorem{ex}{Example}
\newcommand{\add}[1]{\textcolor{black}{#1}}

\pagestyle{plain}

%

%

%

\title{AutoCE: An Accurate and Efficient Model Advisor for Learned Cardinality Estimation}

%
%
%

\author{\IEEEauthorblockN{Jintao Zhang$^1$, Chao Zhang$^1$, Guoliang Li$^1$, Chengliang Chai$^2$}
	\IEEEauthorblockA{\textit{$^1$Department of Computer Science, Tsinghua University, $^2$School of Computer Science, Beijing Institute of Technology} \\
		\{zjt21@mails., cycchao@mail., liguoliang@\}tsinghua.edu.cn, ccl@bit.edu.cn
\thanks{Guoliang Li and Chao Zhang are the corresponding authors. This work is supported by NSF of China (62232009, 61925205, 62102215, 62072261), Huawei, TAL education, and Beijing National
Research Center for Information Science and Technology.}
}
}

%
%
%
%
%
%
%
%
%
%
%
%
%
%
%
%
%
%

\maketitle

\begin{abstract}
Cardinality estimation (CE) plays a crucial role in many database-related tasks such as query generation, cost estimation, and join ordering. Lately, we have witnessed the emergence of numerous learned CE models. However, no single CE model is invincible when it comes to the datasets with various data distributions. To facilitate data-intensive applications with accurate and efficient cardinality estimation, it is important to have an approach that can judiciously and efficiently select the most suitable CE model for an arbitrary dataset.

In this paper, we study a new problem of selecting the best CE models for a variety of datasets. This problem is rather challenging as it is hard to capture the relationship from various datasets to the performance of disparate models. \add{ To address this problem, we propose a model advisor, named \autoce, which can adaptively select the best model for a dataset. The main contribution of \autoce is the learning-based model selection, where deep metric learning is used to learn a recommendation model and incremental learning is proposed to reduce the labeling overhead and improve the model robustness.} 
We have integrated \oursys into PostgreSQL and evaluated its impact on query optimization. The results showed that \oursys achieved the best performance (27\% better) and outperformed the baselines concerning accuracy (2.1x better) and efficacy (4.2x better).
\end{abstract}
%
%

\section{Introduction}~\label{sec:intro}

Machine learning (ML) based CE models \cite{cidr2019/mscn, pvldb/DuttWNKNC19, vldb2020/deepdb, wu2020bayescard, pvldb/naru2019, pvldb/LearnedCost, pvldb/YangKLLDCS20, wu2021unified, pvldb/FACE2022,pvldb/iris, conf/ssdbm/KDE, ICDE2022-ML4AL,journals/pvldb/ML4DB21, sigmod22/HTAP-Database-Survey, li2021ai} have recently attracted significant attention because it can harness the strong learning and representation ability of ML models to achieve superior performance~\cite{ sun2021learned,wang2020we,vldb2022/alibaba_learnedEvaluation, sigmod/LCE-evaluation-POSTECH}. There are various types of learned CE models that either (1) encode the query workload to model the relationship between queries and their cardinalities (\textit{query-driven} approaches~\cite{cidr2019/mscn, pvldb/DuttWNKNC19,pvldb/LearnedCost}), or (2) encode the datasets to model the joint data distribution (\textit{data-driven} approaches~\cite{vldb2020/deepdb, wu2020bayescard, pvldb/naru2019, pvldb/YangKLLDCS20, pvldb/FACE2022}).

\begin{figure}[!t]
\centering
\includegraphics[width=0.49\textwidth]{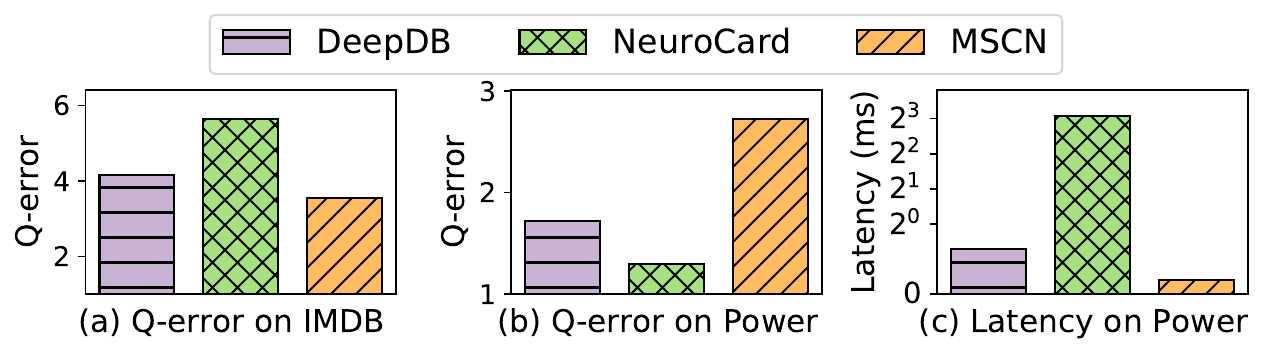}
\vspace{-2.5 em}
\caption{Experiment of CE models over different datasets.}
\vspace*{-2.3 em}
\label{fig:motivation}
\end{figure}

\begin{ex}[Motivation]
Figure \ref{fig:motivation} shows an experimental study of several typical learned CE models on different datasets. Figure~\ref{fig:motivation}(a) reports the estimation accuracy of 3 methods (DeepDB \cite{vldb2020/deepdb}, NeuroCard \cite{pvldb/YangKLLDCS20} and MSCN \cite{cidr2019/mscn} on \texttt{IMDB} dataset. Figure~\ref{fig:motivation}(b) reports the accuracy of these methods on a different \texttt{Power} dataset. Figure~\ref{fig:motivation}(c) reports the inference efficiency of the CE models.
\end{ex}

Given the results, we have the first observation that on dataset \texttt{IMDB}, the accuracy is MSCN $>$  DeepDB  $>$ Neurocard, while on \texttt{Power}, it is Neurocard $>$  DeepDB  $>$ MSCN. The estimation accuracy varies because datasets have diverse and complicated data characteristics~\cite{sun2021learned,wang2020we}, e.g., the number of tables, the number of joins, data correlation and skewness. There is no single CE model to perform well across the entire feature space. For example, MSCN performs better on \texttt{IMDB} mainly because the dataset contains multiple tables and it is hard for the other two data-driven methods to model the joint distribution across tables. Although we can train different CE models, test the performance and select the best one, it is rather time-consuming~\cite{Sagemaker_whitebox}, which is not practical when datasets are diverse and evolving. Hence, given a dataset, it is necessary to develop a  model advisor that can well capture the various data features and identify the best CE model precisely and quickly. In terms of the inference efficiency, MSCN performs the best, followed by DeepDB, and Neurocard is the most inefficient one.  Given the \texttt{Power} dataset, if a  user aims at generating millions of benchmarking queries with  cardinality constraints~\cite{sigmod22/LearnedSQLGen, UniBench, UniBench_journal, DBLP:conf/vldb/Zhang18}, the CE step of the generator needs to be efficient, so she is likely to choose MSCN. If she expects a good trade-off, DeepDB may be chosen rather than the most accurate Neurocard. 

\noindent \textbf{Applications.} Overall, in this paper, we aim to build an intelligent CE model advisor that \textit{given any particular dataset, it can efficiently choose the most suitable CE model customized to the user's requirements, e.g., the estimation accuracy and efficiency.} Many applications can benefit from the model advisor. A representative is the cloud data services \cite{sigmod/snowflake, CloudDatabases}, which provide elastic data storage and querying services for multiple tenants. Using the model advisor, the cloud vendors can select an accurate CE model for an arbitrary dataset without the costly online learning. Particularly, the model advisor can rapidly select a new model when any data drift is detected~\cite{Sagemaker_monitor}. Moreover, the model advisor can facilitate many applications that consider both accuracy and efficiency of CE models such as fraud detection~\cite{cikm/account_detection} and query generation~\cite{sigmod22/LearnedSQLGen}.

\noindent \textbf{Challenges.} An ideal solution is to take the given dataset and the user's requirement as input, then directly predict the best CE model among all candidates. There are \add{two} main challenges for learning such an advisor. (C1) the feature space of real-world datasets is huge, thus it is challenging to well capture the relationship from various datasets to the performance of different CE models. (C2) labeling the datasets is prohibitively expensive as it requires training and testing all the candidate CE models against each dataset to get the performance as the label.

To address the above challenges, we propose \autoce, a model advisor for learned cardinality estimation. To well capture the relationship from data features to the CE models' performance, we model the features of a dataset as a graph and leverage deep metric learning to train a similarity-aware encoder of datasets based on the performance of CE models (addressing C1). To reduce the labeling overhead, we collect the samples that are not well predicted in the validation phase, then augment those samples with well-predicted samples by combining their features and labels (addressing C2).

\begin{figure}[!t]
\centering
\includegraphics[width=0.49\textwidth]{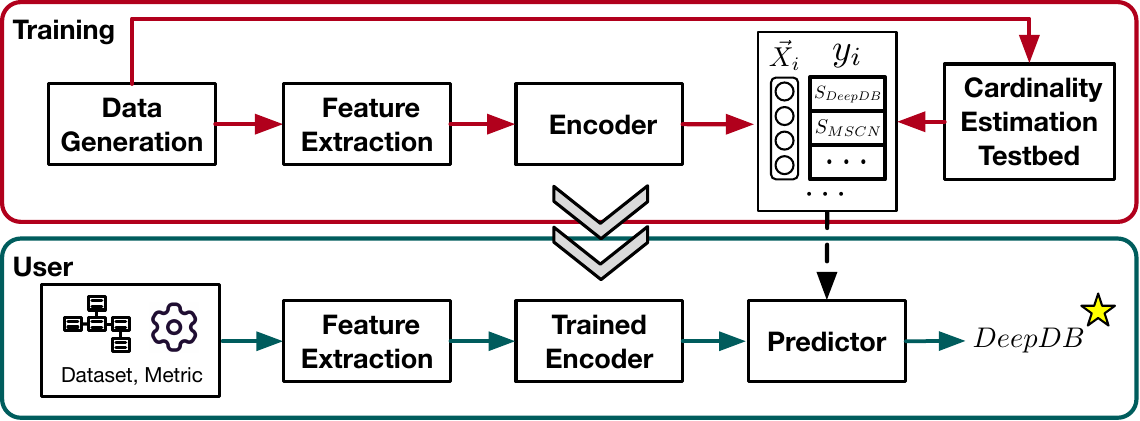}
\vspace{-1.8em}
\caption{An example of AutoCE: after the offline training, it selects a tailored cardinality estimation model with an arbitrary dataset and specified metrics.}
\vspace{-1.5em}
\label{main_workflow}
\end{figure}

To summarize, we have made the following contributions:

\begin{enumerate}[leftmargin=*]
\setlength\itemsep{0em}
    \item We study a new problem of selecting the learned CE models for a set of datasets. We propose a model advisor, \autoce, which takes any dataset and metrics as input, then quickly selects a tailored CE model.

    \item \add{We propose a new learning method to encode the datasets as feature graphs, and leverage the deep metric learning to train a similarity-aware graph encoder}.

    \item \add{We develop an incremental learning phase that identifies poorly-predicted samples and synthesizes new training samples to train the model incrementally.}
    
    \item We have integrated \oursys into PostgreSQL v13.1 by injecting the estimated cardinalities into its query optimizer. The experimental results showed that \oursys significantly improved the query performance by 27\%, and improved the accuracy and efficacy by 2.1x and 4.2x, respectively.

\end{enumerate}

\section{Problem Statement}

\noindent  \textbf{CE-model selection problem.} Given a set of learned cardinality estimation (CE) models $\mathbb{M}=\{M_{1},.,M_{m}\}$, and a set of datasets $ \mathbb{D} =\{D_{1},.,D_{n}\}$,  the CE-model selection problem aims to select an optimal CE model $M_i \in \mathbb{M}$ for each dataset $D_j \in \mathbb{D}$ based on the specified performance metrics. The widely-used metrics include:

\noindent [1. \textit{Q-error}] is an accuracy metric~\cite{moerkotte2009preventing} for evaluating a selected model, defined as  $\textit{Q-error} = \frac{max(\widehat{card}, card)}{min(\widehat{card},card)}$, where $\widehat{card}$ is estimated cardinality of a query and $card$ is the ground truth.

\noindent [2. \textit{Inference latency}] is the efficiency metric for evaluating the inference overhead of a selected model, which is the running time of using the model for cardinality estimation. 

\noindent [3. \textit{E2E latency}] is used to denote the end-to-end (E2E) latency of answering a query using the selected CE model.

We quantify the Q-error/inference latency/E2E latency of a model $M_i$ on a dataset $D_j$ with three steps. First, we execute the given testing queries to get the true cardinalities (if not given, we generate the testing queries against $D_j$). Second, we use model $M_i$ to estimate the cardinalities for the testing queries, then obtain the Q-error and inference latency. Third, we inject the estimated cardinalities to the database, then measure the E2E latency.

\begin{ex}[A Working Example of AutoCE]
As shown in Figure \ref{main_workflow}, \oursys consists of an offline training phase and an online prediction phase (i.e., without online learning). In the training phase, it generates a variety of datasets, extracts the dataset features, and learns an encoder to fit the relation from the data features to the performance of CE models. Particularly, the CE models are evaluated using the developed CE testbed, and the encoder is trained with different combinations of performance metrics. In the prediction phase, it takes as input any dataset and metric weights (e.g., [0.5, 0.5] represents 50\% Q-error and 50\% inference latency), then extracts its features and obtains the embedding using the trained encoder, and finally uses a predictor to select the best CE model.

\end{ex}

\begin{figure*}[!t]
\centering
\includegraphics[width=0.95\textwidth]{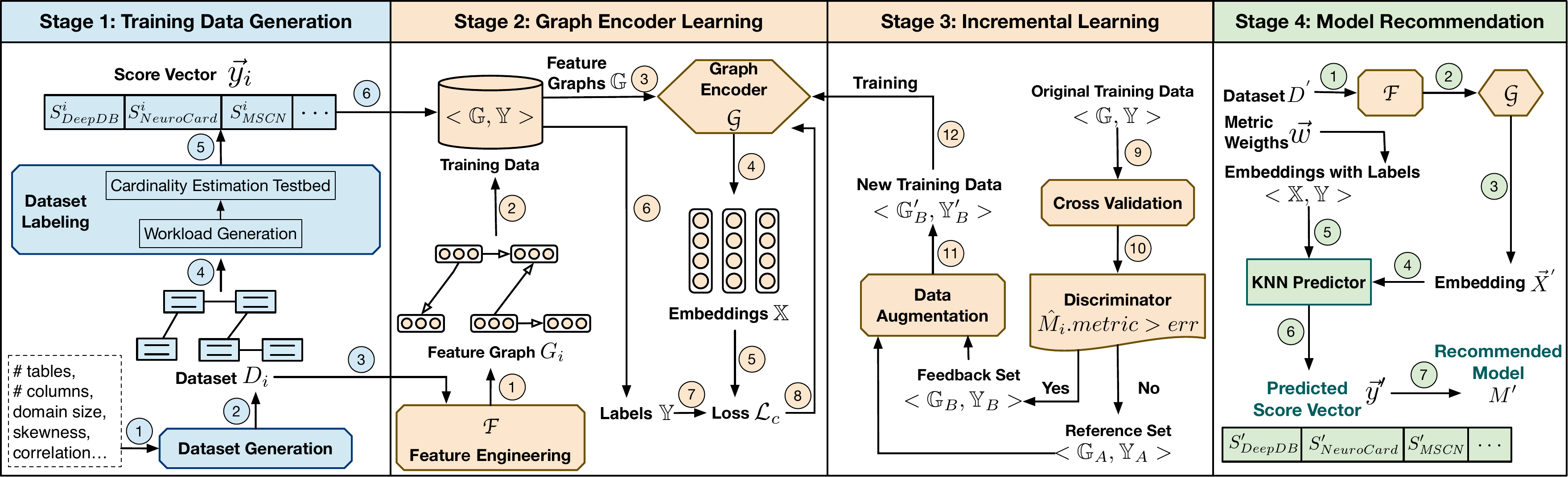}
\vspace{-1em}
\caption{\add{An overview of AutoCE on model recommendation including data preparation (Stage 1), training (Stage 2-3) and recommendation (Stage 4).}}

\vspace{-1.5em}
\label{sys_overview}
\end{figure*}

\vspace{1mm}

\vspace{-5pt}
\section{Overview of AutoCE}
\vspace{-2pt}
\label{sec:Overview}
An overview of \oursys is presented in Figure \ref{sys_overview}, which is divided into four stages. \add{Stage 1 is for data preparation (\scircled{1}-\scircled{6}). The training phase consists of stages 2 and 3 by walking through steps \scircled{1}-\circled{12}. Particularly, stage 2 performs the deep metric learning to train a graph encoder, and stage 3 conducts the incremental learning based on the trained graph encoder and the original training data. Finally, stage 4 makes the recommendation with steps \scircled{1}-\scircled{7} }.

\noindent \textbf{Stage 1 \add{(Data Preparation)}:} \add{\scircled{1}-\scircled{3}} \oursys takes as input a set of parameters (such as the number of tables/columns, domain size, skewness, and correlation), then generates a number of datasets. \add{\scircled{4}-\scircled{6}}After that, each dataset is labeled by a unified cardinality estimation testbed, and each label is a score vector that measures the performance of the CE models. 

\noindent \textbf{Stage 2 \add{(Training Phase I)}:} \add{\scircled{1}-\scircled{2}} \oursys extracts the CE-related features for each dataset through a process of feature engineering, and models the extracted features as a feature graph. To learn a graph encoder that can produce similarity-aware dataset embeddings w.r.t. the performance of the CE models, \add{\scircled{3}-\scircled{8}} \oursys conducts the deep metric learning by recursively learning from batches of labeled feature graphs with loss back-propagation.

\noindent \textbf{Stage 3 \add{(Training Phase II)}:} \add{\scircled{9}-\circled{10}} \oursys validates the performance of the trained graph encoder with all the training data. If certain feature graphs are not well-predicted, \add{\circled{11}} \oursys augments them with well-predicted feature graphs by combining their features and labels. Then, \add{\circled{12}} \oursys conducts incremental learning with the new training data to improve the generality of the advisor.

\noindent \textbf{Stage 4 \add{(Recommendation)}:} \oursys makes the recommendation for a target dataset and specified weights of metrics. Specifically, \add{\scircled{1}-\scircled{3}} it goes through the process of feature engineering, acquires the embedding of the target dataset with the updated encoder, \add{\scircled{4}-\scircled{5}} and then searches for the $k$-nearest neighbors based on its distances to the embeddings of labeled datasets. Finally, \add{\scircled{6}-\scircled{7}} an averaged score vector is computed based on the neighbors' labels associated with the metric weights, and the top ranker of the score vector corresponds to the selected model.

\subsection{Training Data Generation}
\label{sec:overview-DP}

\noindent \textbf{(1) Dataset Generation.} As existing datasets are hard to cover diverse features for CE, we present the \textit{dataset generation} component, which can provide thousands of datasets to cover a relatively comprehensive space of data features. Particularly, it takes a set of input parameters, and then generates a set of synthetic tables with diverse distributions of data features. \add{Theoretically speaking, it is rather hard to generate all meaningful data distributions. Nevertheless, \oursys enables a good generalization regarding different datasets as it generates the data based on selected data features for cardinality estimation. To handle the unexpected data distributions, we design an online adaptive method which can detect the unexpected data distribution, then uses online learning to obtain the ground truth, and finally updates the model accordingly. }

\noindent \textbf{(2) Dataset Labeling.} This component is an offline procedure that leverages a unified testbed to obtain the performance (e.g., Q-error and inference latency) of the CE models on each dataset efficiently. We normalize these performance values, and then combine them into a score vector (see Subsection \ref{subsec: normalization}). For each combination of weighted performance metrics, each dataset $D_i$ corresponds to a score vector $\vec y_i$, which has a length of $m$ that indicates the performance of CE models $\mathbb{M}$. Note that the framework of \oursys is extensible, so any newly-emerged CE model and specified metrics can be readily incorporated into the framework.

\subsection{Graph Encoder Learning}

\label{sec:overview-LE}
\noindent \textbf{(1) Feature Engineering.} This component extracts the relevant data features, then represents them as feature graphs. Specifically, it extracts various features of columns of a single table, such as the skewness and cross-column correlations. Meanwhile, it also considers the join correlation across tables. Since a relational schema can be naturally represented as a graph where the nodes correspond to tables and edges correspond to join relations, we represent the extracted features of a dataset as a feature graph. In a feature graph, each node contains the extracted features of a single table, and each edge denotes a join between two tables. Consequently, the feature graphs $\mathbb{G}$ become the input of the graph encoder, and they will be transformed to the dataset embeddings $\mathbb{X}$.

\noindent \textbf{(2) Learning the graph encoder using Deep Metric Learning (DML).} \oursys aims to learn a graph encoder to produce a dataset embedding that can capture the relation between diverse data features and the performance of different CE models. As discussed in Section~\ref{sec:intro}, it is a challenging problem as it is hard to learn the mapping from the datasets to the CE model's performance due to the large datasets-to-performance space. To tackle this problem, we define the similarity between datasets regarding the CE model's performance ($\mathbb{Y}$), then we leverage the high-level idea of deep metric learning that captures performance similarities/dissimilarities between the datasets. To be specific, given a dataset, it pushes together its embeddings $x$ and its similar counterparts ($\mathbb{X}^+$), and pushes apart the embeddings $x$ and its dissimilar counterparts ($\mathbb{X}^-$). The objective to learn an encoder such that $d(f(x),f(x^+))>>d(f(x),f(x^-))$, where $(x, x^+)$ is a positive pair and $(x, x^-)$ is a negative pair, and  $d(\cdot)$ can be the Euclidean distance. Since the number of dataset pairs is much larger than the number of datasets, our DML-based learning approach can well capture the relation from the datasets to the performance of CE models.

\subsection{Incremental Learning}
\label{sec:overview-DA}
\add{The incremental learning is an integral part of \oursys and is used for co-training the recommendation model. The main purpose of incremental learning is to reduce the labeling overhead by augmenting the training data, thereby improving the robustness of the recommendation model.} If the discriminator detects a poorly-predicated sample, it augments them with a well-predicted sample by combining their features and labels. Afterward, the graph encoder is incrementally trained with the new training data. Particularly, it conducts a cross-validation phase with the original training data $<\mathbb{G}$,$\mathbb{Y}>$. For a training sample $<G_i$,$\Vec{y}_i>$, if \oursys recommends a CE model that produces a large error, (i.e., $\hat{M}_i.metric > err$), the discriminator assigns it to the feedback set $<\mathbb{G}_B$, $\mathbb{Y}_B>$. Otherwise, the sample is put to the reference set $<\mathbb{G}_A$, $\mathbb{Y}_A>$. For data augmentation, \oursys generates a new feature graph $G'_i$ with a synthetic label $\Vec{y}'_i$ for each sample $G_i \in \mathbb{G}_B$. Specifically, it first searches for a nearest neighbor $G_j \in \mathbb{G}_A$ based on the Euclidean distance, then generates new training data $<G'_i$,$\Vec{y}'_i>$ by combining the pair of feature graphs $<G_i, G_j>$ and pair of labels $<\Vec{y}_i$,$\Vec{y}_j>$. Consequently, \oursys uses the new training data $<\mathbb{G}'_B$,$\mathbb{Y}'_B>$ to learn an updated graph encoder incrementally.

\subsection{Model Recommendation}
\label{sec:overview-Prediction}

To select a model for a given dataset and specified metric weights $\vec{w}$, we develop a KNN-based predictor. Specifically, for a given new dataset $D'$, \oursys represents the datasets with feature extractor $\mathcal{F}$, utilizes the trained graph encoder $\mathcal{G}$ to output a similarity-aware embedding $\Vec{X}'$, and feeds the embedding to the KNN predictor. The predictor searches for the $k$-nearest embeddings $\{\Vec{X}_{1},.,\Vec{X}_{k}\}$ for the embedding $\Vec{X}'$ based on Euclidean distance, then averages their labels $\{\Vec{y}_{1},.,\Vec{y}_{k}\}$ to a unified score vector $\Vec{y}'$. Finally, the model with the highest  score in vector $\Vec{y}'$ is the selected model $M'$.

\begin{figure*}[!t]
\centering
\includegraphics[width=0.91\textwidth]{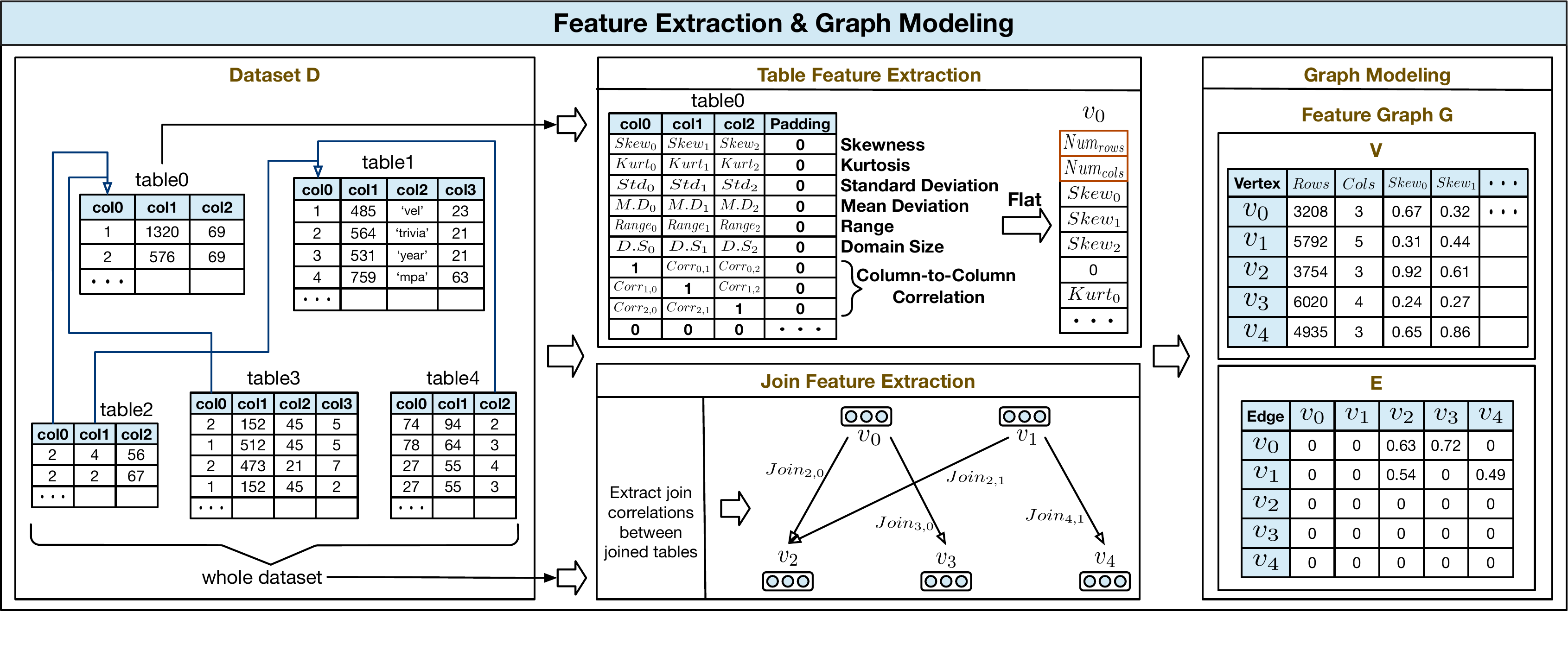}
\vspace{-2.4 em}
\caption{Feature engineering for a dataset D, including the processes of feature extraction and graph modeling.}
\vspace{-1.5 em}
\label{fig:feature extraction}
\end{figure*}

\section{Dataset Generation and Labeling}
\label{sec:data preparation}

\subsection{Dataset Generation}\label{sec:data generation}
The data generation method is divided into two parts: single-table generation and multi-table generation. We generate the datasets based on three data features (i.e., skewness, column correlation, and join correlation) as follows:

\noindent \textbf{(F1) Skewness.} 
The data for each column is generated using the Pareto distribution, which generates a set of random numbers with skewness. We change the skewness parameter $skew$ from 0 to 1, where $skew=0$ means uniform distribution. As the $skew$ increases, the data becomes a more skewed distribution. Specifically, the probability density function (PDF) of the column skewness is defined as follows:

\vspace{-1 em}

\begin{equation}
\label{equ:skewness PDF}
f(x)= \frac { (1+x \cdot (skew-1))^{-1- \frac{-1}{skew-1}} }{v_{max}-v_{min}}
\end{equation}

\noindent where $skew$ is the skewness of a column, $v_{max}$ is the maximum value, and $v_{min}$ is the minimum value.

\noindent \textbf{(F2) Column Correlation.} 
We change the correlation parameter $r$ to control the probability that two columns have the same value in the same position. That is, take two values $(v_1, v_2)$ at the same position in the two columns, and make them equal with the probability of $r$. The larger the probability is, the stronger the correlation is. When the corr is 0, there is no correlation between two columns.

\noindent \textbf{(F3) Join Correlation.} We generate the join correlation for each PK-FK join by varying the probability $p$, Particularly, we generate a join correlation $p$ within the range [$j_{min}$, $j_{max}$], then take a portion of $p$ without replacement from the PK column of the \textit{main table}. After that, we populate the FK column by randomly sampling values from the portion of data. As a result, the higher value of $p$, the larger portion of PK column data is occupied in the FK column data. 

\subsubsection{Single Table Generation}
The generation of a single table takes input as these variables: the number of columns $n$, the number of rows $k$, domain size $d$, skewness parameter $skew$, max correlation $r$, then outputs a generated table with $n$ columns. It works in two steps. First, it sets the max value $v_{max}$ to $d$ and sets $v_{min}$ to 1, selects a $skew$ value within [0,1], then generates the column with $k$ rows based on Equation \ref{equ:skewness PDF}. Second, it iterates over all the generated columns to add a correlation between the two selected columns. Particularly, for every two adjacent columns, we correct their correlation $r$ to a random value between 0 and 1 by modifying their values.

\subsubsection{Multi-Table Generation} Generating multiple tables is based on the single-table generation. \oursys takes as input the table size $n$, the max and min join correlation $j_{max}$ and $j_{min}$, then generates a correlated dataset with $n$ tables in three steps. Firstly, it generates $n$ tables independently using the single-table generation. Secondly, it selects $m$ tables as \textit{main tables}, and assigns a primary key to each main table. Thirdly, it randomly correlates an arbitrary table (could be a main table) with a main table by generating a PK-FK join with a correlation $p$ following the procedure of \texttt{F3}.

\subsection{Dataset Labeling}  \label{sec:Dataset_Scoring}

\subsubsection{Model Training and Testing} We develop a CE testbed to measure the performance of CE models on each generated dataset. We implemented seven state-of-the-art CE models, including three query-driven methods, three data-driven methods, and one hybrid approach.The labeling process consists of four steps. First, it generates a query workload against the dataset. Second, it acquires the true cardinalities by running the queries in the database. Third, it trains the candidate CE models. For data-driven models, they are trained with the dataset directly to learn a joint distribution. For query-driven models, they are trained with a set of encoded training queries with true cardinalities. Finally, it measures their performance score with a set of testing queries and true cardinalities. 

\add{To incorporate a new cardinality estimation baseline into \oursys, we deploy the baseline to the cardinality estimation testbed, which conducts the dataset labeling and produces the corresponding score vectors. Finally, \oursys is able to select the suitable CE model based on the new score vectors.}

\subsubsection{Score Normalization} \label{subsec: normalization} We use $Q \text{-}error$ \cite{moerkotte2009preventing} and inference latency $T$ to measure the models' performance on accuracy and efficiency. We use the mean of $Q \text{-}error$, denoted as $Q \text{-}error_{mean}$ to represent the error of a certain CE model on the testing queries against a dataset. Note that it is possible to use other percentiles of the metrics, such as 50-th, 95-th, and 99-th of $Q \text{-}error$. In this work, we choose the mean as the metric. We also denote $T _{mean}$ as the average estimation time of a model on the testing queries. We normalize the scores with different weights of $Q \text{-}error_{mean}$ and $T_{mean}$ to consider various combinations of them. We have evaluated the improvement of the combinations of $Q \text{-}error$ and $T$ on the E2E latency (See Section \ref{sec:exp_E2E}).

Let $S^{D_i,M_j}$ denotes the performance score of the $j$-th CE model $M_j$ on the $i$-th dataset $D_i$, which is a combination of the accuracy score and the efficiency score. Let $w_a$ (resp. $w_e$) be an accuracy weight (resp. efficiency weight), where $w_a+w_e=1$ and $0 \leq w_a \leq 1$ with a step of 0.1. The score $S^{D_i,M_j}$ is defined as follows:

\vspace{-1 em}
\begin{equation}
\label{equ:optimal_score}
S^{D_i,M_j}  = 	w_a * S^{D_i,M_j}_{a} + w_e * S^{D_i,M_j}_{e}   
\end{equation}

\noindent where  $S_{a}^{D_i,M_j}$ and $S_{e}^{D_i,M_j}$ denote its normalized accuracy score and efficiency score, respectively. Specifically, their corresponding formulas are follows:

\vspace{-1 em}
\begin{equation}
\label{equ:accuracy score}
S_{a}^{D_i,M_j} = \frac{Max(Q\text{-}error_{mean}^{D_i})-Q\text{-}error_{mean}^{D_i,M_j}}{Max(Q\text{-}error_{mean}^{D_i})-Min(Q\text{-}error_{mean}^{D_i})}
\end{equation}

\begin{equation}
\label{equ:efficiency score}
S_{e}^{D_i,M_j} = \frac{Max(T_{mean}^{D_i})-T_{mean}^{D_i,M_j}}{Max(T_{mean}^{D_i})-Min(T_{mean}^{D_i})}
\end{equation}

In the above two formulas, $Max(Q\text{-}error_{mean}^{D_i})$ and $Max(T_{mean}^{D_i})$ represent the maximum $Q\text{-}error_{mean}$ and $T_{mean}$ of all cardinality estimation models $\mathbb{M}$ against dataset $D_i$; $Q\text{-}error_{mean}^{D_i,M_j}$ and $T_{mean}^{D_i,M_j}$ represents the $Q\text{-}error_{mean}$ and $T_{mean}$ of $j$-th cardinality estimation method $M_j$ against the $i \text{-}th$ dataset.

\begin{definition}
\label{def:D-error}
\textit{(D-error).} We define a metric, called $D\text{-}error$, based on the performance store $S^{D_i,M}$. The rationale of $D\text{-}error$ is to measure how far the performance score of $M$ is from that of the optimal model $M_{opt}$ regarding dataset $D_i$. The optimal model $M_{opt}$ refers to the model with the highest score on the dataset. The formula is denoted as: $D\text{-}error=\frac{S^{D_i,M_{opt}}-S^{D_i,M}}{S^{D_i,M}}$, where $S^{D_i,M}$ is the performance score of model $M$ on dataset $D_i$. 
\end{definition}

\section{Model Training and Inference}
\label{sec: learned embedding}

\subsection{Feature Engineering} \label{sec:feature extraction}
\subsubsection{Feature Extraction} \label{sec:joincorrelation}

Conventionally, the machine learning techniques encode the tuples, train a model, and do the inference for a tuple. In our task, a training sample is a dataset instead of a tuple, and \oursys does the inference for a dataset each time.
Therefore, we only need to capture the data distributions that are relevant to the overall performance of the cardinality estimation models. \add{We extract three relevant features (i.e., column correlation, skewness, and domain size) for each column. We also use other related data features as complements, including kurtosis, range, and standard/mean deviation, as well as the number of rows and columns.} We also consider the features of joins for multi-table datasets. Instead of using the join selectivity that often has a small value, we compute the join correlation by taking the set of the FK column data of a table, then calculating its ratio over the PK column data of a joined table. By doing so, the network model can better capture the join correlation. The extraction of skewness, column correlation, and join correlation is a reverse process of the data generation (See three processes (F1), (F2), and (F3) in Section \ref{sec:data generation}).

\subsubsection{Graph Modeling}
After the features of a dataset have been extracted, we model them as a graph, called \textit{feature graph}, in which the vertices of the graph correspond to the features of the table, and the edges of the graph correspond to joins across the tables; the weight on an edge is computed based on the join correlation between two tables. A feature graph consists of a \textit{vertex matrix} and an \textit{edge matrix}, and graph modeling involves vertex modeling and edge modeling.

\noindent \textbf{(1) Vertex Modeling.}
For a dataset, we set the maximum number of columns of all tables to $m$. \add{For each table, there are two unique features: the number of rows and the number of columns.} For each column of a table, we extract $k$ data features. For every two columns of a table, we extract a total of $m \times m$ correlation features. In total, a table can have a maximum of $(k+m)\times m+2$ features. If a generated table has less than $m$ columns, we use padding to fill the empty positions with 0. Consequently, a $n$-table dataset has a vertex matrix $V$ with the shape of V is $[n,(k+m)\times m+2]$.

\noindent \textbf{(2) Edge Modeling.} Suppose a dataset has $n$ tables, an edge matrix $E$ has the size of $n \times n$. If there is an FK in the $j\text{-}th$ table referencing the PK of the $i\text{-}th$ table, then the value of $E[i][j]$ stores the join correlation, otherwise $E[i][j]=0$.

\begin{ex}
As shown in Figure \ref{fig:feature extraction}, given a five-table dataset $D$, we extract features from each table with a maximum of 4 columns. The vertex features of table $i$ are flattened as a vector $v_i$, i.e., $v_0$ for $table_0$. The vertex matrix $V$ has the shape of $[5,(6+4)\times 4+2]=[5,42]$, and the edge matrix $E$ has a shape of $5 \times 5$ and contains four non-empty join weights. For instance, the FK column of table $v_2$ has a portion of 54\% of the PK column of table $v_1$.
\end{ex}

\subsection{Encoding the Feature Graph} \label{sec:GIN}

We encode the feature graphs using Graph Isomorphism Network (GIN)\cite{xu2018powerful}, which is a state-of-the-art graph neural network. GIN conducts nonlinear mapping of the feature graph through $L$ \glayer layers. For \glayer layer, GIN multiplies the features of each vertex $v$’ neighbors and the corresponding edge features between $v$ and its neighbors, and then each vertex $v$ aggregates the multiplication result, producing a representation of the vertex. Finally, GIN takes as input the representations of all vertices and outputs a representation using sum pooling.

Specifically, our graph encoder consists of $L$ \glayer layers and one layer of sum pooling. It takes a \textit{feature graph} $G=(V,E)$ as input and outputs an encoded vector $\vec X$. We denote $H^l$ as the output of the $l\text{-}th$ \glayer layer, where $H^0$ is the input of the first \glayer layer with ($H^0=V$). Let $h^l_i$ denote the feature vector of the $i\text{-}th$ vertex output by the $l\text{-}th$ \glayer layer, which is computed by:

\vspace{-7 pt}
\begin{equation}
\label{equ:Ginconv}
    h_i^{(l+1)} = f_{\theta} \left( (1+\epsilon)h_i^l + \sum\nolimits_{j \in N(i)} e_{ji}^{'} \cdot h_j^l \right)
\end{equation}
\vspace{-7 pt}

In the above formula, $f_{\theta}$ represents a function determined by a learnable parameter $\theta$ in the network; $\epsilon$ is also a learnable parameter; $N(i)$ represents the joined neighbors of vertex $i$, and $e_{ji}^{'}$ represents the join correlation between vertex $j$ and $i$.

\subsection{Learning the Graph Encoder} \label{sec:Contrastive Learning}

\begin{figure}[!t]
\centering
\includegraphics[width=0.492\textwidth]{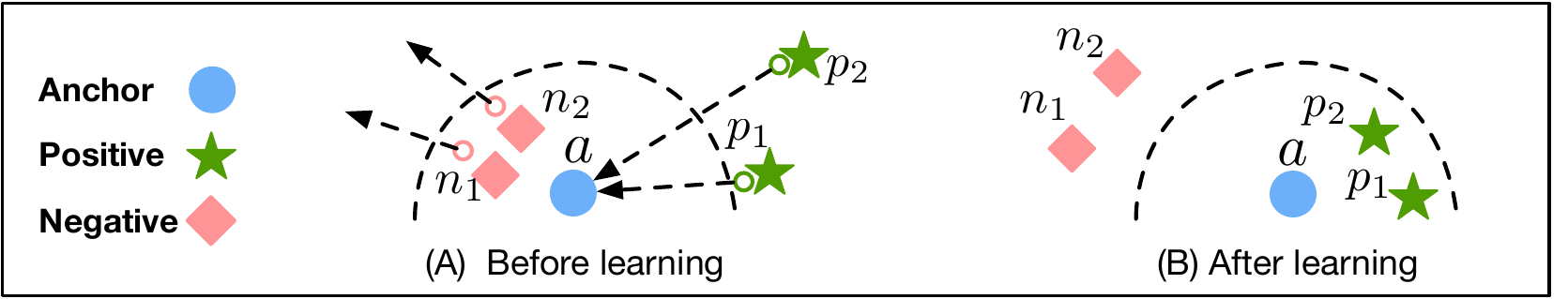}
\vspace{-1.7em}
\caption{Learning effect of graph contrastive learning.} 
\vspace{-.5em}
\label{fig:learningEffect}
\end{figure}

\begin{figure}[!t]
\centering
\includegraphics[width=0.487\textwidth]{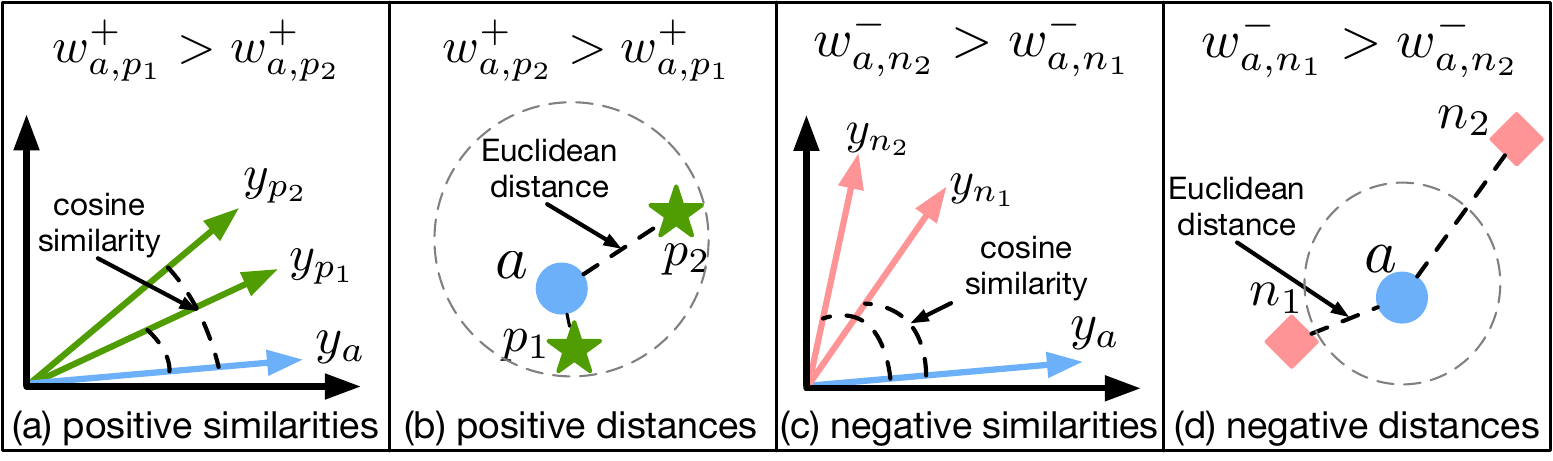}
\vspace{-1.7em}
\caption{Pair weighting: $w^+_{a,p1}$ and $w^-_{a,n1}$ denotes the weight for a positive pair $<a,p1>$ and negative pair $<a,n1>$, respectively.}
\label{fig:pairWeighting}
\vspace{-1.5em}
\end{figure}

We learn the graph encoder by deep metric learning. Particularly, we treat the score vectors as labels, and use the labels to determine whether two feature graphs are in the same classes or not (See Eq.~\ref{equ:similarity} and Eq.~\ref{equ:p or n pair}). Intuitively, the closer two score vectors are, the more likely two feature graphs are in the same class. Since we have the score vectors with various combinations (weights of $w_a$ and $w_e$), our graph encoder can learn from each combination of score vectors to support various users' requirements (See Eq.~\ref{equ:optimal_score}).

\begin{ex}Figure~\ref{fig:learningEffect} illustrates the learning effect of deep metric learning. There are three types of graph instances: anchor, positive, and negative instances. Each instance is an embedding representation for a feature graph. After the learning process, the positive instances \{$p_1$, $p_2$\} have been pulled closer to the anchor while the negative instances \{$n_1$, $n_2$\} have been pushed away from the anchor $a$.
\end{ex}

\begin{figure}[!t]
\centering
\includegraphics[width=0.49\textwidth]{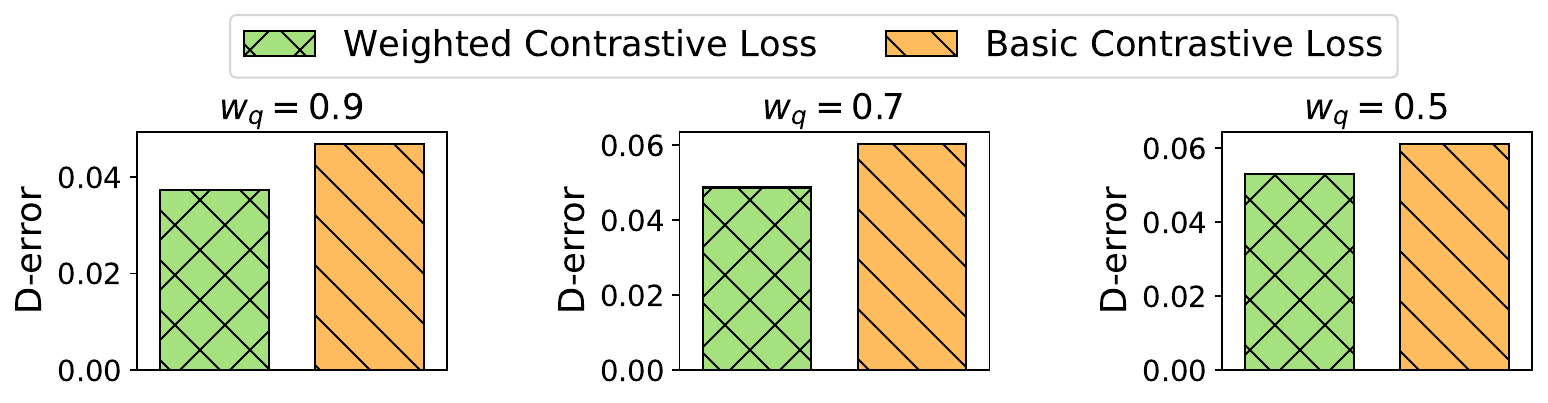}
\vspace{-2 em}
\caption{Comparisons between two contrastive loss functions.}
\label{fig:loss_func}
\vspace{-1.5 em}
\end{figure}

\noindent \textbf{Measuring the similarity with score vectors.} We use score vectors to decide whether two instances belong to the same class or not.  We first define the similarity and label as follows:

\begin{definition}
\textit{(Performance Similarity).} Given the labels $\vec y_i$ and $\vec y_j$ of \textit{feature graphs} $G_i$ and $G_j$, We use cosine similarity to define their performance similarity $\mathit{Sim}_{ij}$ as follows:

\vspace{-.5 em}
\begin{equation}
\label{equ:similarity}
\mathit{Sim}_{ij}= \frac{\vec y_i \cdot \vec y_j}{||\vec y_i|| \ ||\vec y_j||}
\end{equation}
\end{definition}

\begin{definition}
\textit{(Positive/Negative Instance).} Given a pair of feature graphs $<G_i$, $G_i>$ and a threshold $\tau$, we denote the $G_i$ as an anchor, then $j$ belongs to either the positive index set $\mathcal{P}_{i}$ or the negative index set $\mathcal{N}_{i}$ based on their similarity $S_{ij}$ and threshold $\tau$. Particularly, $G_j$ is a positive instance of $G_i$ if $S_{ij}$ is greater than $\tau$. Otherwise, $G_j$ is a negative instance of $G_i$. The formula is defined as follows:

\vspace{-.5 em}
\begin{equation}
\label{equ:p or n pair}
    G_j.index \in \left\{
	\begin{aligned}
	\mathcal{P}_{i} & \quad\quad\quad \mathit{Sim}_{ij} \geq \tau \quad\\
	\mathcal{N}_{i} & \quad\quad\quad \mathit{Sim}_{ij} < \tau \quad\\
	\end{aligned}
	\right
	.
\end{equation}
\end{definition}

We define $U_{i,j}$ as the distance between the embeddings of the two \textit{feature graphs} $G_i$ and $G_j$. We use the Euclidean distance function to define the $U_{i,j}$ as follows:

\vspace{-.5 em}

\begin{equation}
\label{equ:D_w}
    U_{ij} = ||\vec{X_i} - \vec{X_j}||_2
\end{equation}

\noindent where $\vec{X_i}$ and $\vec{X_j}$ are the embeddings of feature graph $G_i$ and feature graph $G_j$, respectively.

\noindent \textbf{Weighted contrastive loss.} We define a new loss function, called weighted contrastive loss, to take into account both the distance weight and similarity weight. To the best of our knowledge, no existing works~\cite{simbad/tripleloss, cvpr/liftloss, cvpr/multiloss} have considered the weight for label similarity.  Intuitively, since we wish to capture the differences between graph pairs based on their similarities to an anchor, we can (1) assign larger weights to the positive pairs with larger similarities/distances, and (2) allocate larger weights to the negative pairs with smaller similarities/distances.

\begin{ex}
Figure~\ref{fig:pairWeighting} illustrates four cases of pair weighting for either performance similarities or embedding distances. For case (a) where two positive instances $p_1$ and $p_2$ have different similarities, the larger the similarity ($Sim_{a,p1}>Sim_{a,p2}$), the more important the larger pair ($w^+_{a,p1}>w^+_{a,p2}$). For case (b) where two positive instances $p_1$ and $p_2$ have different distances, the larger the distance ($U_{a,p2}>U_{a,p1}$), the more important the larger pair ($w^+_{a,p2}>w^+_{a,p1}$). On the contrary, the negative pairs are more important if they have smaller similarities/distances since we would like to push them apart. For case (c), the smaller the similarity, the larger the weight ($w^-_{a,n2}>w^-_{a,n1}$). For case (d), the smaller the distance, the larger the weight ($w^-_{a,n1}>w^-_{a,n2}$).
\label{exa:pairWeighting}
\end{ex}

Apart from the consideration of absolute similarity weight and distance weight, we also consider the relative weight. The intuition is similar to the Example \ref{exa:pairWeighting} but using the relative weight: for a positive (resp. negative) pair $<G_i, G_j>$, the larger (resp. smaller) the relative similarity/distance, the more (resp. less) important the pair $<P_i, P_j>$ is. For instance, for a negative pair $<G_i, G_j>$ with a set of neighbors $K$, the smaller the relative similarity ($Sim_{ij}-Sim_{ik}$), the more important the pair $<G_i, G_j>$ is. By putting it all together, we define the weighted contrastive loss as follows:

\begin{definition}
\textit{(Weighted Contrastive Loss).}
Given a batch of $m$ \textit{feature graphs}, the loss $\mathcal L_c$ is defined as follows:

\vspace{-1.2 em}

\footnotesize

\begin{equation}
\label{equ:our Contrastive Loss m}
    \mathcal{L}_c = \frac{1}{m} \sum_{i=1}^{m} \left (log  \sum_{k \in \mathcal{P}_{i}} \left(e^{ U_{ik}+\mathit{Sim}_{ik} } \right) + log \sum_{k \in \mathcal{N}_{i}} \left(e^{ (\gamma -U_{ik}-\mathit{Sim}_{ik}) } \right) \right)
\end{equation}
\normalsize

\noindent where $\mathcal{P}_{i}$ and $\mathcal{N}_{i}$ are the positive index set and the negative index set of the anchor $G_i$; $\gamma$ is a fixed margin.
\end{definition}

To verify the effectiveness of the loss function in (9), we compare our weighted contrastive loss function with the basic contrastive loss~\cite{hadsell2006dimensionality} as follows:

\vspace{-1 em}

\begin{equation}
\small
\label{equ:basic Contrastive Loss m}
\mathcal{L} = \frac{1}{m} \sum_{i=1}^{m} \left (\sum_{k \in \mathcal{P}_{i}} U_{ik} - \sum_{k \in \mathcal{N}_{i}} U_{ik} \right)
\end{equation}

As shown in Figure~\ref{fig:loss_func}, the new loss function produces a better result than that of the basic contrastive loss on 200 synthetic datasets.. The main reason is that our loss function considers both distance weight and similarity weight while the basic contrastive loss neglects such information.

\begin{algorithm}[!t] 
\small
\caption{DML-based Graph Encoder Learning} 
\label{alg:Graph Contrastive} 
\KwIn{All $n$ feature graphs with labels, batch\_size $m$.}
\KwOut{A trained GIN $\mathcal{G}$.}
Initialize GIN network parameters $\theta$ \;
\For{each epoch in training process}{
    \For{each batch in this epoch}{
        \For{$i \gets 1$ \textbf{to} $m$} {
            \For{$j \gets 1$ \textbf{to} $m$}{
                Calculate $\mathit{Sim}_{i,j}$ according to Eq. \ref{equ:similarity} \;
                    \If{$\mathit{Sim}_{i,j} \geq \tau$}{
                        $\mathcal{P}_i.add(j)$ ; \tcp{Positive index}
                    }
                    \Else{
                         $\mathcal{N}_i.add(j)$ ; \tcp{Negative index}
                    }
            }
            $\vec{X_i} = \mathcal{G}(G_i)$ \tcp{Embedding with Eq.\ref{equ:Ginconv}}
        }
    
        Calculate $\mathcal{L}_c$ according to Eq. \ref{equ:our Contrastive Loss m} \;
        $\theta = \theta - \eta \cdot \nabla\mathcal{L}_c$ ; \tcp{learning rate $\eta$}
    }
}
\textbf{return} $\mathcal{G}$
\end{algorithm}
\setlength{\textfloatsep}{1mm}

\noindent \textbf{Analysis of pair weighting.} The optimization goal of the loss function is to shorten the embedding distance of the positive pairs, and enlarge the embedding distance of the negative pairs. By differentiating $L_c$ on $U_{ij}$, then we compute the weights for the positive and negative pairs as follows:

\vspace{-1.2 em}
\begin{equation}
\label{equ:positive weight}
    w_{i,j}^+ = \left| \frac {\partial L_c} {\partial U_{ij} } \right| = \frac{1} { \sum_{k \in \mathcal{P}_i } e^{ [(U_{ik}-U_{ij})+(\mathit{Sim}_{ik}-\mathit{Sim}_{ij})] } } \quad
\end{equation}
\vspace{-.5 em}
\begin{equation}
\label{equ:negetive weight}
    w_{i,j}^- = \left| \frac {\partial L_c} {\partial U_{ij} } \right| = \frac{1} { \sum_{k \in \mathcal{N}_i } e^{[(U_{ij}-U_{ik})+(\mathit{Sim}_{ij}-\mathit{Sim}_{ik})] } } \quad
\end{equation}

\noindent where $w_{i,j}^+$ and $w_{i,j}^-$ denotes the weight for a positive pair and a negative pair, respectively. It is not difficult to observe that, as the relative distance ($U_{ij}-U_{ik}$) and the relative similarity ($\mathit{Sim}_{ij}-\mathit{Sim}_{ik}$) increase, the positive pairs have larger weights, and the negative pairs have smaller weights.

\noindent \textbf{Description of the training algorithm.} Algorithm \ref{alg:Graph Contrastive} illustrates the procedures of the similarity-aware graph encoder learning. We initialize the GIN network parameters $\theta$ (Line 1). For each epoch (Line 2), a set of $n$ feature graphs are divided into $\lceil \frac{n}{m} \rceil$ batches. The training steps for each batch are as follows. First, it computes the similarity $\mathit{Sim}_{i,j}$ of each graph pair $<G_i$,$G_j>$ (Line 3-6). For each anchor $G_i$, if $\mathit{Sim}_{i,j}$ is greater than a threshold $\tau$, the index of the instance $G_j$ is assigned to the positive index set $\mathcal{P}_{i}$ (Line 7-8). Otherwise, it is added to the negative index set $\mathcal{N}_{i}$ (Line 9-10). Second, for each instance $G_i$, the embedding $\vec{X_i}$ is generated based on Eq.~\ref{equ:Ginconv} (Line 11). Then, it calculates the weighted contrastive loss with the entire batch according to Eq.~\ref{equ:our Contrastive Loss m} (Line 12). Third, it updates the parameters $\theta$ of GIN $\mathcal{G}$ via the backpropagation of the loss $\mathcal{L}_c$ (Line 13). When all epochs have been conducted, it returns the trained GIN $\mathcal{G}$  (Line 14).

\subsection{Recommendation using Learned Encoder} \label{sec:prediction}

\noindent \textbf{A KNN-based predictor.} With a trained GIN $\mathcal{G}$, we can acquire an embedding for an arbitrary dataset. The key idea is to compare the distances between the labeled embeddings to the unlabeled target dataset's embedding, then find the k-nearest neighbors' labels associated with the users' requirements. We define the recommendation candidate set as follows:

\begin{definition}
\label{RCS}
\textit{(Recommendation Candidate Set (RCS)).} Given a trained GIN network $\mathcal{G}$ and $n$ feature graphs $\mathbb{G}$ with labels $\mathbb{Y}$, the recommendation candidate set is $(\mathbb X, \mathbb Y)$ which contains all the embeddings of $\mathbb{G}$ with labels $\mathbb{Y}$.

\end{definition}

\noindent \textbf{Recommendation process.} \oursys makes the recommendation in three steps. First, it feeds all $n$ training feature graphs into the trained GIN $\mathcal{G}$ to generate their embeddings $\mathbb X$. Since these feature graphs have been labeled in the data preparation phase, their labels $\mathbb Y$ are also available. Hence, the \textit{RCS} $(\mathbb X, \mathbb Y)$ is obtained. Second, \oursys takes a target set $D'$ as input, produces its feature graph $G'$, and generates its embedding $\vec X^{'}$ with the trained GIN $\mathcal{G}$. After that, it searches for its k-nearest neighbors $\{\vec X_1,\dots,\vec X_k\}$ in the \textit{RCS} based on their embeddings $\mathbb X$'s distances to the target embedding $\vec X^{'}$. Then the k-nearest neighbors' indexes are added to the set $\mathbb{K}$. Third, \oursys averages the labels associated with $\mathbb{K}$ into a score vector, and outputs the top ranker's corresponding model as the recommended model. The formula is given as follows:

\vspace{-1em}
\begin{equation}
    \label{equ:recommend}
    M' = \max (\frac {\sum_{i=1}^{k} \vec {y}_{\mathbb{K}[i]} } {k}).index
\end{equation}
\vspace{-1em}

\noindent where $\mathbb{K}$ is the indexes of the k-nearest neighbors, and $\vec y$ is a vector of the corresponding labels; $M'$ is the recommended model that has a max score in the score vector.

\subsection{Online Adapting for Unexpected Data Distribution}
\add{To handle the unexpected data distribution, we design an online adaptive method: it detects the data distributions that are out of the trained distributions. Then, it uses the online learning to select the CE model, and finally it updates the model accordingly. More specifically, the online adaptive method works in three steps. First, given a target dataset, it detects if there is any data drift, i.e., a data distribution is disparate from the existing RCS based on the feature graph's Euclidean distance. We set a distance threshold to determine if a dataset is apart from the RCS. Particularly, the distance from a dataset $D$ to the RCS is defined as the closest distance from $D$ to all the feature graphs in RCS. We sort the distance values of the feature graphs in RCS in ascending order, then take the 90th percentile distance value as the threshold. If the distance of a dataset is greater than the threshold, it is considered as an unexpected data distribution. Second, for an unexpected dataset, it employs the online learning to get the ground truth, then adds the new labeled sample to the RCS. Third, it uses the new training sample to update the recommendation model by deep metric learning.}

\section{Incremental Learning}
\label{sec: data augmentation}

\begin{algorithm}[!t]  
\small
\caption{Incremental Learning with Mixup}
\label{alg:Incremental Learning} 
\KwIn{Original training data $<\mathbb{G}$, $\mathbb{Y}>$, threshold $b$, GIN $\mathcal{G}$.}
\KwOut{An updated GIN $\mathcal{G}$.}

%
%
%
%
%
%
%
%
$\mathbb X = \mathcal{G}(\mathbb{G})$ ; \tcp{Get all Embeddings}
Split $<\mathbb{G}, \mathbb{Y}, \mathbb X>$ to $<\mathbb{G}_1,\mathbb{Y}_1,\mathbb X_1>, ... ,<\mathbb{G}_\xi,\mathbb{Y}_\xi,\mathbb X_\xi>$ \;
%
%
\For {$v$ \textbf{in} $[1,\xi ]$} { 
    Get a validation set $(\mathbb X_v, \mathbb{Y}_v)$ \;
    $\mathbb X_c, \mathbb{Y}_c$ = $\mathbb X \setminus \mathbb X_v, \mathbb Y \setminus \mathbb Y_c$ ; \tcp{Get RCS}
    \For{$X_i, G_i$ \textbf{in} $\mathbb{X}_v, \mathbb{G}_v$} {
        $\mathbb{K}$ = $\mathit{KNN}(\vec X_i, \mathbb X_c, k)$  \;
        $\hat M_i = Recommend(\mathbb{Y}_{c},\mathbb{K})$ ; \tcp{Eq.\ref{equ:recommend}}
       
        \If{$\hat M_i.D\text{-}error > b$}{
            $\mathbb{G}_B.add(G_i), \mathbb{Y}_B.add(\vec{y}_i)$ ;
        }
        \Else{
            $\mathbb{G}_A.add(G_i), \mathbb{Y}_A.add(\vec{y}_i)$ ; 
        }
    }
}
%

%
%
    
%
%
%
%

\For{$G_i, \vec{y}_i$ \textbf{in} $\mathbb{G}_B, \mathbb{Y}_B$} {
    Get the nearest neighbor $G_j$ and $\vec{y}_j$ in $\mathbb{G}_A,\mathbb{Y}_A$ \;
    $G_i', \vec y_i' = Mixup(G_i, \vec{y}_i, G_j, \vec{y}_j)$ ; \tcp{Eq.\ref{equ:dataset augment} }
    $\mathbb{G}'_B.add(G_i), \mathbb{Y}'_B.add(\vec{y}_i)$ ;
            \tcp{Get new sample}
}

Incrementally train a $\mathcal{G}$ according to Algorithm \ref{alg:Graph Contrastive}  \;
\textbf{return} $\mathcal{G}$ \;
\end{algorithm}

\subsection{Incremental Learning with Data Augmentation}

%

%

\noindent \textbf{Incremental learning process.} The incremental learning process has three steps: (1) collect the feedback and reference; (2) augment the data; and (3) train a new encoder. We introduce the procedures in Algorithm \ref{alg:Incremental Learning} with the following steps:

\noindent \textbf{Step1.} \oursys takes as input the training data $<\mathbb{G}$, $\mathbb{Y}>$, a threshold $b$, and a trained GIN $\mathcal{G}$. First, it generates all embeddings $\mathbb{X}$ of $\mathbb{G}$ with GIN $\mathcal{G}$ (Line 1). Then it equally splits the training data with embeddings to $\xi$ folds. Second, it adopts a cross-validation approach to collect the feedback and references (Line 3-12). Specifically, it recursively takes a validation set $(\mathbb X_v, \mathbb{Y}_v)$, and reserves the rest of the folds as the recommendation candidate set (RCS). For each instance $X_v \in \mathbb{X}_v$, it conducts a validation process to collect the feedback and reference by the KNN-based prediction (Line 7), it produces a recommended model $\hat M_i$ (Line 8). If the model $\hat M_i$ has a D-error greater than the threshold $b$, then its feature graph and label are added to the feedback set $<\mathbb{G}_B, \mathbb{Y}_B>$ (Line 9-10). Otherwise, it goes to the reference set $<\mathbb{G}_A, \mathbb{Y}_A>$ (Line 11-12).

%

%

%

\noindent \textbf{Step2.}
For each sample $<G_i$, $\Vec{y}_i>$ in the feedback set $<\mathbb{G}_B, \mathbb{Y}_B>$, \oursys uses \textit{Mixup} to generate a new sample (Line 13-16). Specifically, it finds the nearest neighbor $<G_j$, $\Vec{y}_j>$ in the reference set based on the Euclidean distance. Then it generates a new sample $<G'_i$, $\Vec{y}'_i>$ by employing \textit{Mixup} to incorporate linear interpolations into their feature vectors and labels, respectively. The formula is as follows:

%

\vspace{-.5em}
\begin{equation}
\label{equ:dataset augment}
\begin{aligned}
    G'_i &= \lambda G_i + (1-\lambda) G_j \\
    \vec y'_j &= \lambda \vec y_i + (1-\lambda) \vec y_j
\end{aligned}
\end{equation}
\vspace{-.5em}

\noindent where $\lambda \sim Beta(\alpha, \beta)$, a Beta distribution within [0, 1].
%
%

\noindent \textbf{Step3.} With the new training samples $<\mathbb{G}_B, \mathbb{Y}_B>$ and the original training data, \oursys incrementally trains a more robust GIN $\mathcal{G}$ (Line 17-18). 

\section{Experiments}
\label{sec:EXperiments}

\vspace{-2pt}
\subsection{Experiment Setting}
\noindent \textbf {Datasets.} As shown in Table \ref{tab:datasets parameter}, we use two real-world datasets and 1,200 synthetic datasets.

\textit{(1) Real-world datasets.} We evaluate \oursys on two real-world datasets: a movie-rating dataset IMDB~\cite{sun2021learned} and a STATS~\cite{vldb2022/alibaba_learnedEvaluation} dataset from the Stack Exchange network. We name them IMDB-light and STATS-light (See Table \ref{tab:datasets parameter}). We use them as unseen testing datasets to verify the effectiveness of \oursys. Since each dataset offers only one testing sample for \autoce, we adopt a split approach to obtain more testing samples. Specifically, we randomly select 20 testing samples from each dataset in two steps: (1) randomly select 1-5 joined tables from the dataset with the join keys; (2) randomly select 1-2 non-key columns for each chosen table. We name them as \imdb and \stats.

\textit{(2) Synthetic datasets.} The generation of synthetic datasets is described in Section \ref{sec:data generation}. We generate 1,200 synthetic datasets where 1000 datasets are used as the training set, and 200 datasets are kept unseen for testing \oursys.

\noindent \textbf {Training and Inference Time.} With all the training data (1000 labeled datasets), \oursys takes 107s to train the graph encoder in an offline manner. For the testing set, the average inference time of \oursys is 0.79s for each dataset. 

\begin{table}[!t]
\centering
\caption{Statistics of Datasets.}
\label{tab:datasets parameter}
\scalebox{0.993}{
    \begin{tabular}{|c|c|c|c|c|}\hline
    Dataset  &  \#Table  &  \#Row  &  \#Column  &  Total domain size \\\hline
    \textbf{IMDB-light}   &   6  &  2.1K-339K  &   12   &   $ 4.3 \times 10^4 $ \\\hline
    \textbf{STATS-light}   &   8   &   1K-328K   &   23   &   $ 1.9 \times 10^5 $ \\\hline
    \textbf{Synthetic}   &   1-5   &   10K-50K   &   2-25   &  $ 1.6 \times 10^4 $ \\\hline
    \end{tabular}
}

\end{table}

\noindent \textbf {Workloads.} We generate 10,000 SPJ queries similar to \cite{pvldb/naru2019, pvldb/YangKLLDCS20}, from which 9,000 queries are used for training each query-driven method and 1,000 queries are kept for testing. We also use the CEB-IMDB benchmark~\cite{marcus2021flow}, where we use all the query templates, but for the ease of implementation, we eliminated the 'GROUP BY' and 'LIKE' predicates in queries.

\noindent \textbf {Metrics.} We use $D\text{-}error$ in Definition \ref{def:D-error} to measure the recommendation error regarding a dataset. Particularly, we use a small $D\text{-}error$ threshold (e.g., 0.1) to indicate if a CE model is accurately recommended or not. We also consider the metrics of Q-error, inference latency, recommendation efficiency, and end-to-end latency.

\noindent \textbf {Environment.} All experiments were performed on a server with a 20-core Intel(R) Xeon(R) 6242R 3.10GHz CPU, an Nvidia Geforce 3090ti GPU, and 256GB DDR4 RAM.

\noindent \textbf {Model Selection Baselines.} We compare with 4 baselines. 

(1) \texttt{MLP-based Selection}. We implement an MLP-based baseline by appending a multilayer perceptron of three layers to the GIN network. This baseline treats the model selection problem as a classification task and trains the recommendation model with the cross-entropy loss \cite{rubinstein1999cross}.

(2) \texttt{Rule-based Selection}. We design a rule-based selection: (i) randomly selecting data-driven models for single-table datasets, and (ii) randomly choosing query-driven models for multi-table datasets.

(3) \texttt{Knn-based Selection}. We develop a Knn-based baseline that extracts the features of datasets, then directly uses the KNN predictor based on the distances of the features rather than the embeddings to select models.

(4) \texttt{Sampling-based Selection}. We implement an online learning baseline. Specifically, it samples a portion of a dataset, trains and tests all the CE models against the samples, and finally selects the best-performing CE model.

\begin{table*}[!t]
	\centering
	\caption{Recommendation accuracy of \oursys and two baselines over 200 synthetic datasets and 40 real-world datasets.}
	\vspace{-.6em} 
	\begin{minipage}{0.47\textwidth}
		\centerline{\textbf{(a)} $w_a=1.0$}
		\scalebox{0.9}{
		\begin{tabular}{c||c||c|c|c}
			\toprule
			Datasets  &  Advisor  &  \makecell[c]{Accuracy\\ \footnotesize ($\epsilon$ = 0.1)}  &  \makecell[c]{Accuracy\\ \footnotesize ($\epsilon$ = 0.15)}  &  \makecell[c]{Accuracy\\ \footnotesize ($\epsilon$ = 0.2)} \\
			\midrule
			\multirow{5}{*}{\texttt{\makecell[c]{Synthetic\\(200)} }}  &  \texttt{MLP-based}  &  56\%  & 66\%  & 74\%  \\
                                                 &  \texttt{Rule-based}  &  16\%  &  23\%  &  31.5\%  \\
                                                 &  \texttt{Knn-based}  &  44\%  & 50\%  &  61\%  \\
                                                 &  \texttt{Sampling}  &  44\%  &  51\%  &  53.5\%  \\
                                                 &  \texttt{AutoCE}  &  \textbf{81.5\%}  &  \textbf{86\%}  &  \textbf{87.5}\%  \\
			\hline
			\multirow{5}{*}{\texttt{IMDB-20}}  &  \texttt{MLP-based}  &  55\%  &  70\% &  75\%  \\
                                               &  \texttt{Rule-based}  &  10\%  & 20\% &  20\%  \\
                                               &  \texttt{Knn-based}  &  40\%  & 40\%  &  50\%  \\
                                                &  \texttt{Sampling}  &  25\%  &  25\%  &  30\%  \\
                                               &  \texttt{AutoCE}  &  \textbf{80\%}  &  \textbf{85\%}  &  \textbf{95}\%   \\
			\hline
			\multirow{5}{*}{\texttt{STATS-20}}  &  \texttt{MLP-based}  &  45\%  &  50\% &  60\%  \\
                                                &  \texttt{Rule-based}  &  5\%  &  10\% &  15\%  \\
                                                &  \texttt{Knn-based}  &  45\%  & 65\%  &  70\%  \\
                                                &  \texttt{Sampling}  &  45\%  &  55\%  &  55\%  \\
                                                &  \texttt{AutoCE}  &  \textbf{85\%}  &  \textbf{85\%}  &  \textbf{90}\%  \\
			\bottomrule
		\end{tabular}
		}
	\end{minipage}
	\hfill
	\begin{minipage}{0.256\textwidth}
		\centerline{\textbf{(b)} $w_a=0.9$}
		\scalebox{0.9}{
		\begin{tabular}{||c|c|c}
			\toprule
			\makecell[c]{Accuracy\\ \footnotesize ($\epsilon$ = 0.1)}  &  \makecell[c]{Accuracy\\ \footnotesize ($\epsilon$ = 0.15)}  &  \makecell[c]{Accuracy\\ \footnotesize ($\epsilon$ = 0.2)} \\
			\midrule
			55\%  &  69.5\%  &  78\%  \\
            20\%  &  28\%  &  37\%  \\
            34\%  &  45.5\%  &  58.5\%  \\
            22.5\%  &  28.5\%  &  37\%  \\
            \textbf{89\%}  &  \textbf{94\%}  &  \textbf{97}\%  \\
			\hline
			50\%  &  75\%  &  80\%  \\
            20\%  &  20\%  &  20\%  \\
            45\%  &  55\%  &  55\%  \\
            15\%  &  20\%  &  25\%  \\
            \textbf{90\%}  &  \textbf{95\%}  &  \textbf{100}\%  \\
			\hline
			40\%  &  60\%  &  65\%  \\
            10\%  &  10\%  &  20\%  \\
            50\%  &  55\%  &  65\%  \\
            30\%  &  40\%  &  45\%  \\
            \textbf{70\%}  &  \textbf{80\%}  &  \textbf{85}\%  \\
			\bottomrule
		\end{tabular}
		}
	\end{minipage}
	\hfill
	\begin{minipage}{0.256\textwidth}
		\centerline{\textbf{(c)} $w_a=0.7$}
		\scalebox{0.9}{
		\begin{tabular}{||c|c|c}
			\toprule
			\makecell[c]{Accuracy\\ \footnotesize ($\epsilon$ = 0.1)}  &  \makecell[c]{Accuracy\\ \footnotesize ($\epsilon$ = 0.15)}  &  \makecell[c]{Accuracy\\ \footnotesize ($\epsilon$ = 0.2)} \\
			\midrule
			45\%  &  60\%  &  72\%  \\
            28\%  &  37.5\%  &  43\%  \\
            37.5\%  &  48.5\%  &  59.5\%  \\
            18.5\%  &  31.5\%  &  41\%  \\
            \textbf{76.5\%}  &  \textbf{88\%}  &  \textbf{96}\%  \\
			\hline
			55\%  &  65\%  &  70\%  \\
            30\%  &  40\%  &  40\%  \\
            50\%  &  60\%  &  80\%  \\
            15\%  &  15\%  &  25\%  \\
            \textbf{80\%}  &  \textbf{90\%}  &  \textbf{100}\%  \\
			\hline
			50\%  &  65\%  &  75\%  \\
            15\%  &  35\%  &  35\%  \\
            50\%  &  70\%  &  75\%  \\
            10\%  &  20\%  &  20\%  \\
            \textbf{85\%}  &  \textbf{90\%}  &  \textbf{90}\%  \\
			\bottomrule
		\end{tabular}
		}
	\end{minipage}
\label{tab:Accuracy}
\end{table*}

\begin{table}[!t]
\centering

\caption{Evaluation of efficacy on CEB Benchmark.} 
\vspace{-.6em}
\label{tab:ceb_res}
\setlength\tabcolsep{9pt}
    \begin{tabular}{c|c|c|c|c}\hline
    Method  &  \texttt{\oursys}	& \texttt{MSCN}	 &  \texttt{LW-NN}  &  \texttt{LW-XGB} \\\hline
    \textbf{D-error}$_{w_a=1.0}$  &  \textbf{0\%}    &	1.30\% &	7.86\%    &	100\%  \\\hline
    \textbf{D-error}$_{w_a=0.9}$  &   \textbf{3.16\%}    &	11.80\%    &	 7.08\%     &  90.76\%  \\\hline
    \textbf{D-error}$_{w_a=0.7}$   &  \textbf{2.51\%}    &	26.29\%    &	5.502\%    &	82.81\%  \\\hline  
    \textbf{D-error}$_{w_a=0.5}$   &  \textbf{1.74\%}    &	40.78\%    &	3.93\%    &	74.87\%  \\\hline 
    \end{tabular}
\end{table}

\begin{figure*}[!t] \vspace{-0.6em}
\centering
\includegraphics[width=0.95\textwidth]{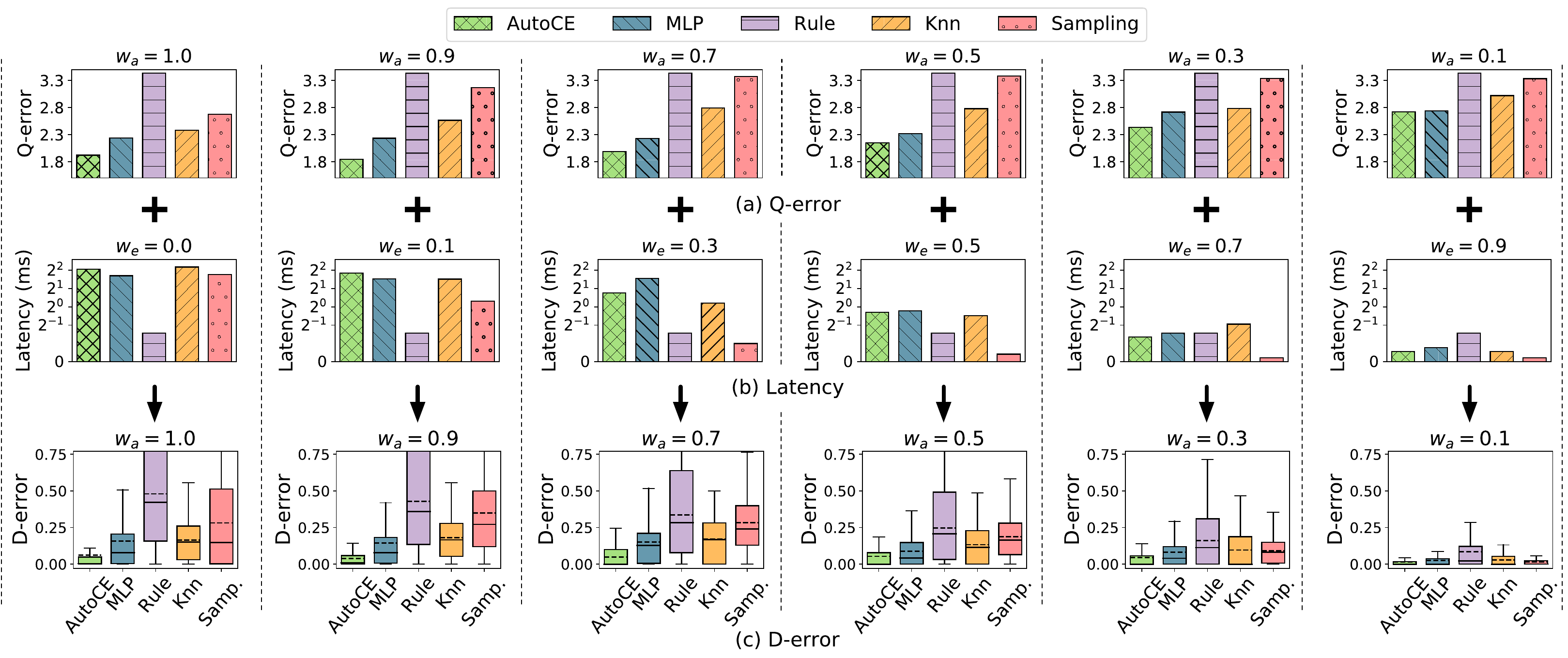}
\vspace{-1em}
\caption{Comparisons between AutoCE and four selection strategies.}
\label{fig:4 strategies}
\vspace{-1.2em} 
\end{figure*}

\noindent \textbf {Cardinality Estimation Baselines.} We compare \oursys with 9 baselines for cardinality estimation as follows: \label{sec:CE methods}

(1) \mscn~\cite{cidr2019/mscn}. This method encodes the query features with a table set, a join set, and a predicate set using one-hot vectors. Then, it utilizes the multi-set convolutional network to learn a mapping function from the feature set to the cardinality.

(2) \xgb~\cite{pvldb/DuttWNKNC19}. It employs a tree-based ensemble method, called XGBoost \cite{chen2016xgboost}, which encodes a query as a sequence of selection ranges, then learns a mapping function from the query encoding to the predicted cardinality.

(3) \nn~\cite{pvldb/DuttWNKNC19}. This approach is based on a lightweight fully connected neural network, and its query encoding method and training strategy are the same as \xgb. 
(4) \deepdb~\cite{vldb2020/deepdb}. This method relies on relational sum-product networks. It divides a table into row clusters and column clusters. Then it utilizes sum nodes (resp. product nodes) to combine the row clusters (resp. column clusters).

(5) \bayes~\cite{wu2020bayescard}. This method relies on Bayes Network to learn a joint distribution.

(6) \neuro~\cite{pvldb/YangKLLDCS20}. It builds a deep autoregressive model on the samples of the full-outer joins of the base tables, then conducts a progressive sampling to make the estimates.

(7) \uae~\cite{wu2021unified}. This method proposes differentiable progressive sampling via the Gumbel-Softmax trick to enable learning from queries. Then it unifies both query and data information using the deep autoregression model~\cite{pvldb/naru2019}.

(8) An ensemble CE method that takes the weighted average estimation of all the CE models (the weight is proportional to their performance on the training datasets).

(9) A default PostgreSQL CE estimator.

\vspace{-6pt}
\subsection{Evaluation of Recommendation Efficacy}

\noindent \textbf{Comparing \oursys with 4 selection baselines.} As shown in Figure~\ref{fig:4 strategies}, we compare the efficacy of \oursys with \mlpb and \ruleb on the synthetic datasets by varying accuracy weights from 1.0 to 0.1. Figure~\ref{fig:4 strategies}(c) shows an overall evaluation of the model performance with D-error, and Figure~\ref{fig:4 strategies}(a) and Figure~\ref{fig:4 strategies}(b) illustrate the breakdown of Q-error and latency, respectively (e.g., the D-error subgraph of $w_a=0.9$ in (c) is broken down to the Q-error subgraph of $w_a=0.9$ in (a) and the Latency subgraph of $w_e=0.1$ in (b)). The results clearly show that \oursys outperforms \mlpb, \ruleb, \sampling and \knn by using deep metric learning and incremental learning. It can be seen that (1) when the accuracy weight is between 0.5 and 1.0, the D-error distribution of \oursys is significantly smaller than that of \mlpb, \ruleb, \sampling and \knn. The mean of D-error of \oursys is 2.5x, 6.7x, 2.8x, and 4.9x better than \mlpb, \ruleb, \sampling and \knn respectively. (2) the mean of Q-error of \oursys is 1.2x, 1.7, 1.3x, and 1.6x better than \mlpb, \ruleb, \sampling, and \knn respectively. (3) when the accuracy weight is between 0 and 0.3, the mean latency of \oursys is 1.6x better than \mlpb. (4) \ruleb has the worst performance with different weights, indicating that the general selection is insufficient to select the CE models for various datasets and performance metrics. (5) \sampling is inferior to \autoce as the performance of CE models has a high variance regarding different samples of datasets. (6) \knn is worse than \autoce because it simply relies on the distances of data features rather than the distances of embeddings that can be aware of the CE models' performance.

\begin{figure}[!t]
\centering
\includegraphics[width=0.475\textwidth]{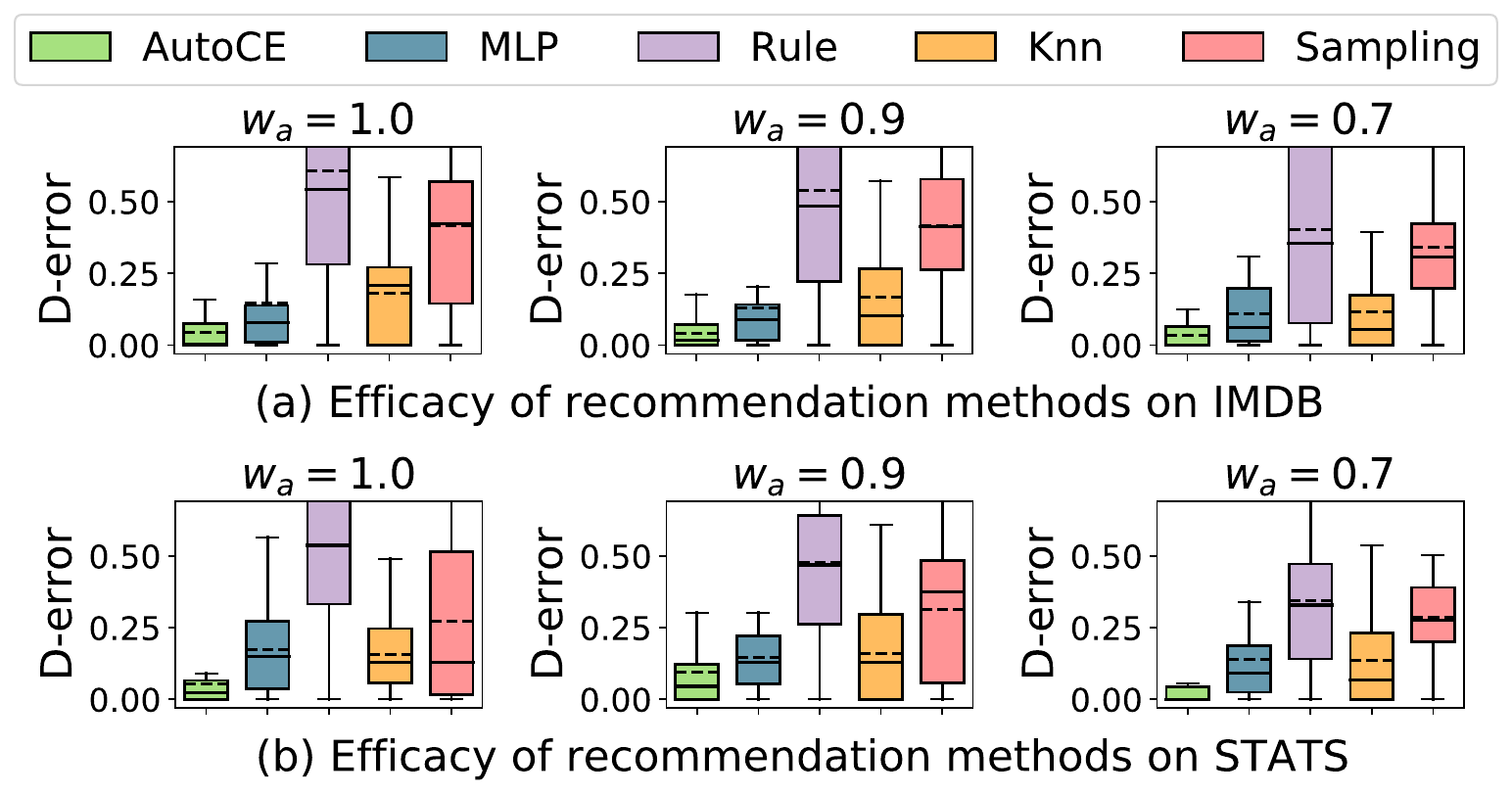}
\vspace{-1.2em}
\caption{Evaluation of efficacy on real-world datasets.}
\vspace{-.2em}
\label{fig:IMDB_STATS}
\end{figure}

For the real-world datasets, the Q-error of \oursys significantly outperforms the other four baselines. As shown in Figure \ref{fig:IMDB_STATS}, we compare the D-error of \oursys with the other four baselines. On \imdb, the mean of D-error of \oursys is 3.2x, 12.7x, 2.9x, and 9.7x better than \mlpb, \ruleb, \sampling, and \knn respectively.  On \stats, the mean of D-error of \oursys is 2.4x, 7.1x, 1.6x, and 4.5x better than \mlpb, \ruleb, \sampling and \knn respectively. 

Regarding CEB-IMDB benchmark, the data-driven cardinality estimators are rather difficult to implement due to the involved large number of tables. For example, 24G GPU memory is not enough for training \texttt{UAE}, and it takes 50 hours to train a \texttt{NeuroCard} model for one dataset. Hence, we just conduct the experiments with the query-driven cardinality estimators including \texttt{MSCN}, \texttt{LW-NN}, and \texttt{LW-XGB}. As shown in Table \ref{tab:ceb_res}, \oursys always achieves the lowest recommendation error with various accuracy weight $w_a$, demonstrating the effectiveness of \oursys. The error of \mscn decreases as $w_a$ decreases, while \nn does the opposite. This is mainly because \mscn is more accurate on most query templates than \nn, but the inference efficiency is not as good as \nn. We found \xgb has a high D-error because it has both a low accuracy and a low inference efficiency.

\begin{table}[!t]
	\centering
	\caption{Autoce's D-error under different $k$.}
	\vspace{-.6em}
	\label{tab:varying k}
	\setlength\tabcolsep{7pt}
		\begin{tabular}{c|c|c|c|c|c}\hline
		$k$  &  1  &  2  &  3  & 4 & 5 \\\hline
		\textbf{D-error}$_{w_a=1.0}$   &   7.67\% 	&   \textbf{6.04\%} 	&   6.74\% 	&   7.01\% 	&   8.71\%   \\\hline
	\textbf{D-error}$_{w_a=0.9}$   &   5.43\% 	&   \textbf{3.73\%} 	&   3.94\% 	&   4.80\% 	&   5.89\%   \\\hline
	\textbf{D-error}$_{w_a=0.7}$   &   6.51\% 	&   \textbf{4.87\%} 	&   5.20\% 	&   5.24\% 	&   6.17\%   \\\hline
	\textbf{D-error}$_{w_a=0.5}$   &   6.62\% 	&   \textbf{5.31\%} 	&   5.89\% 	&   6.37\% 	&   6.94\%   \\\hline
	\end{tabular}
\end{table}

\begin{figure*}[!t]
\centering
\includegraphics[width=1.0\textwidth]{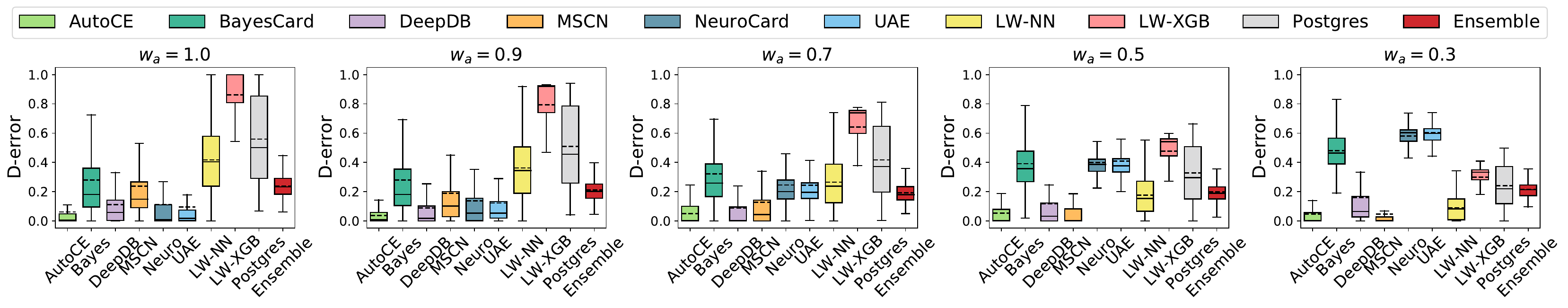}
\vspace{-2.4em}
\caption{Comparisons between AutoCE and seven CE models (plus a PostgreSQL estimator and an ensemble method) on the synthetic datasets.}
\label{fig:Derror of 8methods}
\vspace{-1.6em}
\end{figure*}

\noindent \textbf{Comparing \oursys with 9 CE baselines.} As shown in Figure \ref{fig:Derror of 8methods}, we compare \oursys with 9 CE baselines.  It is clearly visible that the D-error distribution of \oursys is lower than the fixed CE models for different accuracy weight $w_a$. Particularly, the mean of D-error of \oursys is 5.2\% while the average mean of D-error of other fixed CE models is 38.2\%.  In addition, \oursys has 2.8x and 12.3x better model performance than the best-performed one (i.e., \texttt{DeepDB}) and the worst-performed one (i.e., \texttt{LW-XGB}), respectively. These results verify that a fixed CE model or the ensemble method cannot perform well on different types of datasets. What's more, they cannot maintain a lower D-error when the accuracy weight changes. For example, since \uae favors high estimation accuracy by learning from data and queries, it has a higher D-error when $w_a$ decreases. On the contrary, as \nn has a higher estimation efficiency because of the lightweight neural networks, it has a lower D-error as $w_e$ increases. Although the ensemble strategy is relatively stable by combining the CE methods, it has a large margin of D-error to \oursys (20\%).

\noindent \add{\textbf{Comparing \oursys with varying $k$.} We compare the recommendation quality (i.e., D-error) with different $k$. Particularly, we vary $k$ from 1 to 5, and evaluate the recommendation error on synthetic datasets with different accuracy weight $w_a$ from 0.5 to 1.0. As shown in Table \ref{tab:varying k}, D-error is always the smallest when $k = 2$. When $k = 1$, the  performance of the KNN predictor is largely affected by the nearest embedding, which may cause a greater error. When $k \geq 3$, the performance of the predictor is worse because there could be embeddings that are far away from the embedding of the target dataset. }

\vspace{-.3em}
\subsection{Evaluation of Recommendation Accuracy}
\vspace{-.1em}
As shown in Table \ref{tab:Accuracy}, we evaluate the recommendation accuracy of five approaches on the synthetic and real-world datasets. We vary three small threshold values (0.1, 0.15, 0.2) of D-error and three values (1.0, 0.9, 0.7) of the accuracy weight $w_a$. Overall, it clearly shows that \oursys outperforms the four baselines in all settings. On average, the accuracy of \oursys is 1.4x, 2.8x, 1.8x, 2.4x higher than that of \mlpb, \ruleb, \sampling and \knn respectively on the synthetic dataset. For the real-world dataset, the accuracy of \oursys is 1.4x, 4.2x, 1.5x, 3.1x higher than that of \mlpb, \ruleb, \sampling and \knn. We have two observations. First, \oursys can make the recommendation of high accuracy (on average 85\%) on the real-world datasets, which proves that \oursys works on the real-world datasets by using the feature-driven learning method. Second, \oursys outperforms \mlpb with 55\% higher accuracy, indicating the effectiveness of its core parts, i.e., deep metric learning and incremental learning.

\vspace{-.3em}
\subsection{Impact on Query Optimization} \label{sec:exp_E2E}
\vspace{-.1em}
We evaluate the end-to-end latency in PostgreSQL v13.1 with different CE models. Particularly, we use the modified PostgreSQL code of \cite{vldb2022/alibaba_learnedEvaluation} to inject the cardinalities into the query optimizer. To be specific, we invoke each CE model to estimate the cardinalities of all sub-plan queries for a query, then feed them into the query optimizer to generate the query plan. Finally, we execute the query plan in PostgreSQL to get the end-to-end latency. We use 30 synthetic datasets including 15 single-table datasets and 15 multi-table (2-5 tables) datasets, and we run 100 queries against each dataset.

\begin{table}[!t]
	\centering
	\vspace{-.3em}
	\caption{End-to-end Latency in PostgreSQL.}
	\vspace{-.7em}
	\scalebox{0.88}{
	\begin{tabular}{c||c|c|c|c}
		\toprule
		   &  {Single-table}  &  {Multi-table}  &  {Single.}  &  {Multi.} \\
		\midrule
		Method    &  \multicolumn{2}{c|}{Running time + Inference Latency}   &   \multicolumn{2}{c}{Improvement}  \\
		\midrule
		\texttt{PostgreSQL}  &  22.11s + 1.78s   &  1.732h + 16s      &  -  &  -  \\        
        \texttt{TrueCard}   &  \textbf{20.99s}   &  \textbf{1.168h}   &  \textbf{12.14\%}  &  \textbf{32.74\%} \\  
        
		\hline
		\bayes              &  21.68s + 6.78s    &  1.461h + 60s      &   -19.13\%  &  14.90\%  \\  
		\deepdb             &  21.65s + 5.03s    &  1.345h + 36s      &   -11.68\%  &  22.45\%  \\  
		\mscn               &  21.72s + 0.33s    &  1.344h + 3s       &     7.70\%  &  22.55\%  \\ 
		\neuro              &  21.58s + 13.73s   &  1.265h + 125s     &   -47.80\%  &  25.15\%  \\  
		\uae                &  21.66s + 13.07s   &  1.283h + 122s     &   -45.37\%  &  24.16\%  \\ 
		\nn                 &  21.78s + 0.01s    &  1.778h + 0.1s     &     8.80\%  &  -2.40\%  \\ 
		\xgb                &  21.77s + 0.40s    &  1.828h + 4s       &     7.20\%  &  -5.34\%  \\  
		
		\hline
		\autoce$_{w_a=0.5}$  &  \textbf{21.38s + 0.16s}  &  1.319h + 18s            &     \textbf{9.84\%}  &  23.75\%  \\  
		\autoce$_{w_a=1.0}$  &  21.29s + 2.99s           &  \textbf{1.247h + 41s}   &    -1.63\%  &  \textbf{27.56\%}  \\  

		\bottomrule
	\end{tabular}
	}
\label{tab:e2e_time}
\vspace{-.2em}
\end{table}

Table~\ref{tab:e2e_time} shows the total running time of workloads with the model inference latency. For single-table datasets, we find that (1) the estimated cardinalities mainly affect the scan operators of the query plans, e.g., sequential scans or index scans, and (2) the inference latency could be a major factor for the overall execution time. For instance, the inference latency takes more than half of the overall time for \texttt{NeuroCard} and \texttt{UAE} while the query-driven model \texttt{LW-NN} takes only 10ms. \oursys ($w_a=0.5$) achieves the largest improvement ($\sim$9.84\%) as it takes into account both accuracy and efficiency. For multi-table datasets, the plan quality is the major factor for query optimization where the inference latency takes only a small fraction of the overall time. Particularly, more accurate estimated cardinalities result in a better join order and suitable join operators, e.g., hash joins or nested-loop joins. \oursys ($w_a=1$) outperforms  other estimators in performance ($\sim$27.56\%), because \oursys can select the most accurate models for different kinds of datasets.

\vspace{-.3em}
\subsection{Ablation of Deep Metric Learning} \label{sec:exp_gcl ablation}
\vspace{-.1em}
We conduct an ablation study on DML. We implement \autoce (Without DML) by appending three fully connected layers to the GIN network and use the MSE loss function $L= \sum _{i=1}^m ||\vec y_i - \vec {\hat y} ||_2$ to train the the entire network. Then it can recommend a CE model using $max(\vec {\hat y}).index$.

As shown in Figure \ref{fig:ablation}, we compare the D-error of \oursys and \oursys (Without DML) with accuracy weights (0.9, 0.7, 0.5). The result on the synthetic datasets indicates that (1) the D-error of \oursys is always lower than that of \oursys (Without DML) as the accuracy weight $w_a$ decreases. (2) the mean of D-error of \oursys is 40\% better than that of \oursys (Without DML), which verifies the effectiveness of DML.

\vspace{-.4em}
\subsection{Ablation of Incremental Learning} \label{sec:exp_feedback ablation}
\vspace{-.2em}

\begin{figure}[!t]
\centering
\includegraphics[width=0.47\textwidth]{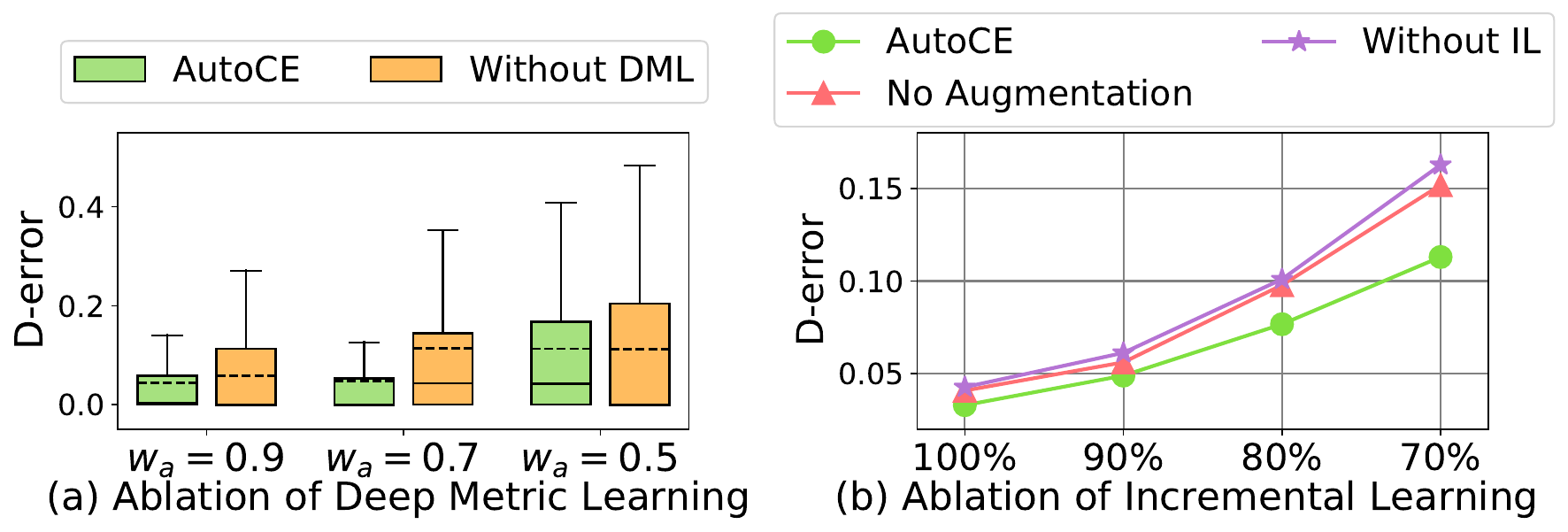}
\vspace{-1em}
\caption{Ablation on core parts of \oursys.}
\vspace{-.3em}
\label{fig:ablation}
\end{figure}

\noindent \textbf{Effectiveness.} In Figure \ref{fig:ablation}, we compare the performance of \oursys with \oursys (Without IL) and \oursys (No Augmentation) using an accuracy weight $w_a=0.9$. The x-axis is the percentage of used training data, and the y-axis is the mean of D-error. We can see that \oursys always has a lower D-error with respect to different proportions of the training data. The main reason is that it can obtain more information through incremental learning. When using 70\% amount of training data, \oursys has $\sim$5\% and $\sim$4\% smaller D-error than that of \oursys (Without IL) and \oursys (No Augmentation).

\noindent \textbf{Efficiency.} Mixup-based incremental learning can synthesize new training samples without labeling. Given 93\% training data, we generate ($\sim$15\%) new training samples with a D-error threshold of 10\% on the feedback, then incrementally train a new encoder with the synthetic data. The results show that it has the same D-error as that of \oursys (Without IL) exploiting 100\% training samples, indicating that \oursys saves about 7\% of the dataset labeling time ($\sim$2 hours).

\begin{figure}[!t]
\centering
\includegraphics[width=0.499\textwidth]{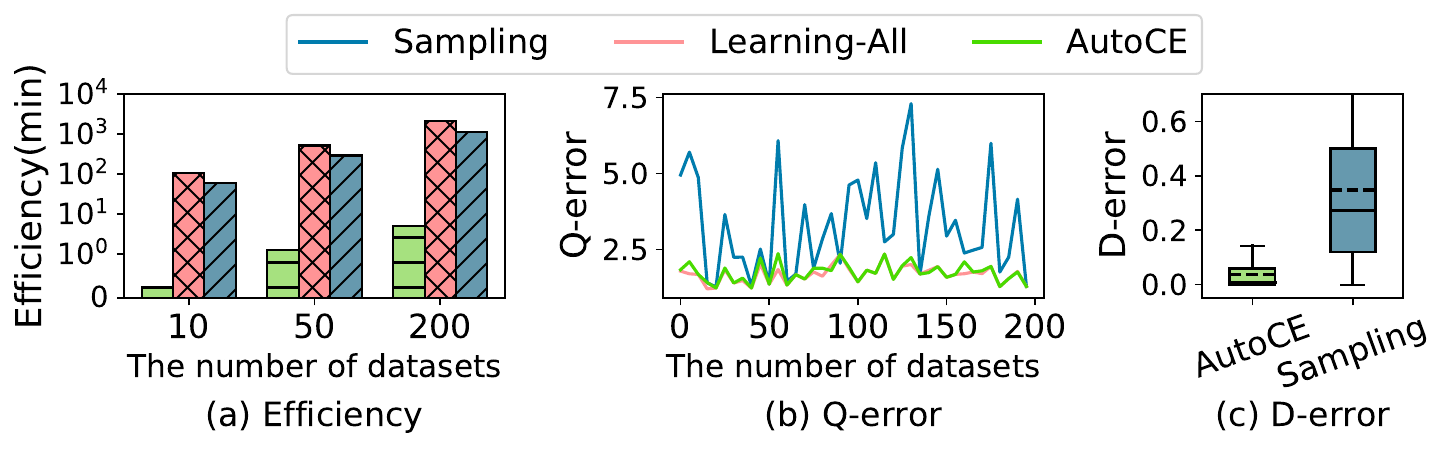}
\vspace{-2.25 em}
\caption{Comparisons between \oursys and online learning methods.}
\label{fig:overhead}
\vspace{-0.3em}
\end{figure}

\begin{figure}[!t]
\centering
\includegraphics[width=0.47\textwidth]{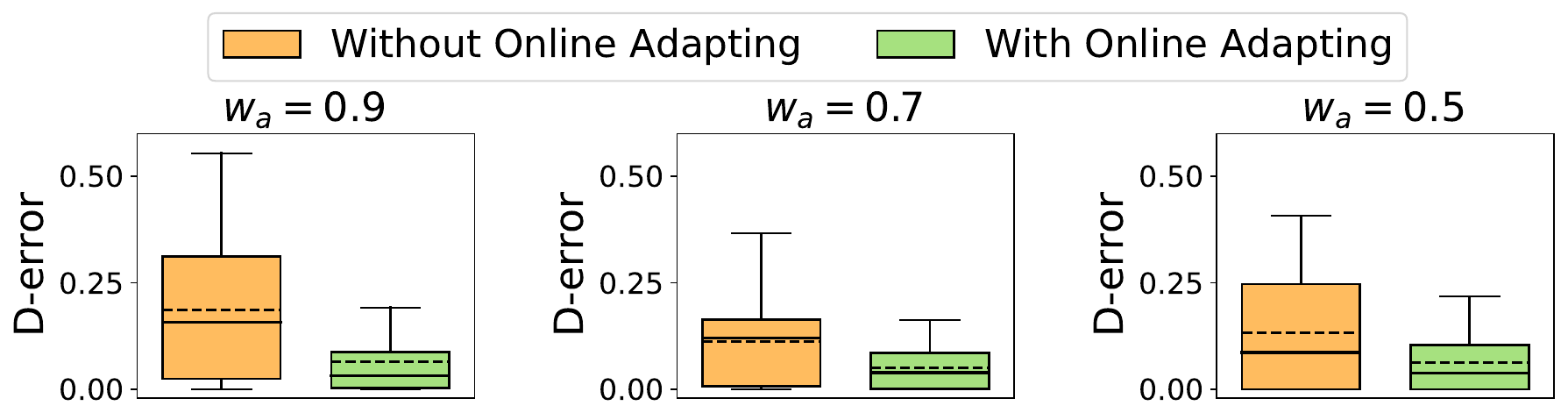}
\vspace{-1 em}
\caption{Ablation of Online Adapting.}
\label{fig:unseen_distrib}
\end{figure}

\vspace{-.3em}
\subsection{Comparing \oursys with Online Learning Methods}
\vspace{-.1em}
We compare \oursys with two online learning methods over 200 datasets. Particularly, the sampling method selects a model by training and testing all the candidate models against a sample of each dataset, and the learning-all (LA) method uses the full dataset each time. As shown in Figure~\ref{fig:overhead}a, two online learning methods incur a significant overhead as the number of datasets increase (LA and sampling approach takes 2230 and 1200 minutes for 200 disparate datasets, respectively), while \oursys requires less than 5 minutes to complete such a task. Figure~\ref{fig:overhead}b depicts that the Q-error of \oursys is close to the LA method (both have a Q-error of 1.8, but the sampling method has a fluctuated Q-error distribution regarding disparate datasets. Figure~\ref{fig:overhead}c shows that the D-error of \oursys is 3.7\% w.r.t. the LA method while the sampling method has a D-error of 34.8\%. In summary, \oursys not only achieves near-optimal CE performance but also reduces the selection overhead significantly (by 455x).

\vspace{-.3em}
\subsection{Ablation of Online Adapting}
\vspace{-.1em}
\add {To verify the effectiveness of the online adaptive approach, we conducted an experiment: (1) we randomly generate datasets and encode them as feature graphs. (2) we deliberately select 100 datasets as the unexpected data distribution based on the distances. (3) we use the 100 datasets for online adapting. We compare the performance of AutoCE with an ablation study on the online adaptive method. As shown in Figure \ref{fig:unseen_distrib}, it can be found that our approach effectively reduces recommendation error by more than 1x on average for the datasets with the unexpected data distributions. }

\section{Related Work}
\noindent \textbf{CE Model Selection.} Several empirical studies \cite{pvldb/CEB21, sun2021learned,wang2020we,vldb2022/alibaba_learnedEvaluation, sigmod/LCE-evaluation-POSTECH} on learned cardinality estimation have been conducted. They analyzed the performance of many ML-based CE models, followed by general rules and insights for selecting the models. For instance, \cite{sun2021learned} suggests query models are more effective for multiple tables, indicating that users could select a query-driven model for multi-table datasets. \cite{vldb2022/alibaba_learnedEvaluation} concludes that only data-driven methods such as FLAT \cite{pvldb/FLAT21}, DeepDB, and BayesCard, can improve the end-to-end performance of the PostgreSQL baseline, meaning that users could select a data-driven model in production. Unfortunately, such general insights are insufficient to select a CE model for diverse datasets. As shown in the experiments, any model could outplay others, depending on datasets and metrics.

\noindent \textbf{Deep Learning Models for Database.} There are currently many works applying deep learning models in databases\cite{zhang2024htap,DBLP:journals/tkde/ZhouCLS22,DBLP:journals/pvldb/YuC0L22,DBLP:conf/sigmod/ZhangCZ022,DBLP:journals/pvldb/0001ZSYHJLWL21,DBLP:journals/pvldb/ZhouSLF20,DBLP:conf/icde/Yu0C020,DBLP:journals/dase/LanBP21,DBLP:journals/dase/WuLZZC22,DBLP:journals/dase/YuanL21,DBLP:journals/dase/PengCX21,DBLP:conf/icde/Yuan0FSH20}. For example, poisoning attack for cardinality estimation, using Monte Carlo tree search to rewrite queries, using graph neural network to generate materialized views, and the hybrid learning-based and cost-based query optimizer~\cite{zhou2021learned, yu2022cost, zhang2024pace}.

\section{Conclusion}
We propose a learned model advisor, \autoce. We leverage deep metric learning to train a graph encoder, which can capture the relation from diverse features of datasets to the performance of cardinality estimation (CE) models. Moreover, we employ the graph encoder and a distance-aware predictor to recommend a learned CE model for a target dataset. We also propose an approach of feedback-driven data augmentation to generate new training samples without labeling the datasets. Experimental results have shown that \oursys significantly outperformed the baselines on both accuracy and efficiency. Moreover, \oursys achieved the best performance compared with other CE models in PostgreSQL.

%
%

\newpage
\balance

%
%
%
%
%
%

\bibliographystyle{abbrv}
\bibliography{main}
\end{document}